\documentclass[conference]{IEEEtran}
\IEEEoverridecommandlockouts
\usepackage{multicol}
\usepackage{multirow}
\usepackage{cite}
\usepackage{amsmath,amssymb,amsfonts}
\usepackage{algorithm}
\usepackage[algo2e,ruled, vlined, linesnumbered]{algorithm2e}
\SetKwComment{kc}{$\triangleright$~}{}
\SetKwComment{tcp}{$\triangleright$~}{}

\SetCommentSty{mycommfont}

\DontPrintSemicolon
\usepackage[pdftex]{graphicx}
\pdfoutput=1
\usepackage{textcomp}
\usepackage{xcolor}
\usepackage{enumitem}
\def\BibTeX{{\rm B\kern-.05em{\sc i\kern-.025em b}\kern-.08em
    T\kern-.1667em\lower.7ex\hbox{E}\kern-.125emX}}
\usepackage{subcaption}
\usepackage{graphicx}
\usepackage{textcomp}
\usepackage{xcolor}
\usepackage{xargs}
\usepackage{xspace}
\usepackage{braket}
\usepackage{bbold}
\usepackage{tikz}
\usetikzlibrary{fit}
\newcommand*\circled[1]{%
  \begin{tikzpicture}[baseline=(1.base), 
      circ/.style={shape=circle,draw,inner sep=1pt}]
    \foreach \num [count=\c,remember=\c as \C (initially 0)] in {#1}{
      \node[circ] (\c) at (\c*3ex,0) {\num};
    }                   
    \ifnum\C>1\node[circ,fit=(1)(\C)]{};\fi
  \end{tikzpicture}%
}

\usepackage{booktabs}
\usepackage{rotating}
\usepackage{makecell}

\definecolor{darkgreen}{rgb}{0,0.5,0}
\definecolor{midnight}{rgb}{0,0.094,0.533}
\definecolor{ocean}{rgb}{0,0.290,0.533}

\usepackage[colorlinks=true,%
   linkcolor=red,%
   citecolor=midnight,%
   urlcolor=blue,%
   bookmarks=false]{hyperref}

\newenvironment{psmallmatrix}
  {\left(\begin{smallmatrix}}
  {\end{smallmatrix}\right)}

\begin{document}

\title{Efficient Hierarchical State Vector Simulation of Quantum Circuits via Acyclic Graph Partitioning}


\author{
\IEEEauthorblockN{Bo Fang\IEEEauthorrefmark{1}\IEEEauthorrefmark{4}, M. Yusuf \"Ozkaya\IEEEauthorrefmark{2}\IEEEauthorrefmark{4},
Ang Li\IEEEauthorrefmark{1},
\"Umit V. \c{C}ataly\"urek\IEEEauthorrefmark{2}
Sriram Krishnamoorthy\IEEEauthorrefmark{1}}
\IEEEauthorblockA{\IEEEauthorrefmark{1} Pacific Northwest National Laboratory, Richland WA, USA }
\IEEEauthorblockA{\IEEEauthorrefmark{2}Georgia Institute of Technology,
School of Computational Science and Engineeering, Atlanta, GA, USA \\
Email: \{bo.fang, ang.li, sriram\}@pnnl.gov, \{myozka, umit\}@gatech.edu }

}


\newcommand{\svsim}{HiSVSIM\xspace}
\newcommand{\dagp}{\texttt{dagP}\xspace}
\newcommand{\dagP}{\texttt{dagP}\xspace}
\newcommand{\nat}{\texttt{Nat}\xspace}
\newcommand{\dfs}{\texttt{DFS}\xspace}
\newcommand{\intel}{IQS\xspace}
\renewcommand{\dblfloatpagefraction}{0.99}


\maketitle

\begingroup\renewcommand\thefootnote{\IEEEauthorrefmark{4}}
\footnotetext{Both authors contributed equally.}
\endgroup
\thispagestyle{plain}
\pagestyle{plain}
\renewcommand{\dblfloatpagefraction}{0.99}

\begin{abstract}
Early but promising results in quantum computing have been enabled by the concurrent development of quantum algorithms, devices, and materials. Classical simulation of quantum programs has enabled the design and analysis of algorithms and implementation strategies targeting current and anticipated quantum device architectures. In this paper, we present a graph-based approach to achieve efficient quantum circuit simulation. Our approach involves partitioning the graph representation of a given quantum circuit into acyclic sub-graphs/circuits that exhibit better data locality. Simulation of each sub-circuit is organized hierarchically, with the iterative construction and simulation of smaller
state vectors, improving overall performance. Also, this partitioning reduces the number of passes through data, improving the total computation time. We present three partitioning strategies and observe that acyclic graph partitioning typically results in the best time-to-solution. In contrast, other strategies reduce the partitioning time at the expense of potentially increased simulation times. Experimental evaluation demonstrates the effectiveness of our approach.
\end{abstract}

\section{Introduction}
\label{sec:intro}
In the last decades, the rapid progress witnessed in the field of quantum
computing involves solid improvement over the quantum algorithms, systems, and
materials. Driven by the scaling complexity of the problems to solve, the
quantum systems would need to respond to the substantially increasing depth of
the operations and the expanding width of the qubit registers. Although the
state-of-the-art quantum computers available to the public now feature 127 qubits~\cite{ibmroadmapupd}, it is still significantly less
than the qubit capacity required by the quantum algorithms. 
Considering the exascale supercomputers capable of up to $10^{17}$ to $10^{18}$
floating-point operations per second (FLOPS), Dalzell et al.~\cite{Dalzell}
calculated that to reach the state of ``quantum supremacy" one needs 208 qubits
with the Instantaneous Quantum Polynomial-time (IQP) circuits~\cite{iqp}, 420
qubits with  Quantum Approximate Optimization Algorithm (QAOA)
circuits~\cite{qaoa} and 98 photons with boson sampling circuits (linear optical
networks)~\cite{boson}. A more common case suggests that there needs to be more
than 2000 qubits and the order of $10^{11}$ of gates with Shor’s
algorithm~\cite{shor} for a practical use. That said, this gap would hinder the
design, implementation, and verification for the quantum algorithm and physical
quantum computing architecture co-design.

Despite the substantial catch-up from the actual quantum hardware effort, several aspects motivate the use of classical systems to simulate the quantum circuit execution. First, designing new quantum algorithms needs iterative studies and trials, where software-based simulators would certainly offer flexible and low-cost platforms. Second, current noisy intermediate-scale quantum (NISQ) technologies usually have a short coherence time, compromising the result correctness of deep circuit execution. Third, publicly available quantum computers (specifically with a large number of qubits, e.g.,~$> 16$) are much less resourceful and usually reside in cloud services, hence the access to those machines is limited. However, as expected, a practical simulation would require a massive memory capacity, thus shifting the quantum circuit simulation to memory-bound and degrading the simulation efficiency.

Since the designated system typically features a large number of compute
nodes with a hierarchical memory architecture to run such simulations, it is
essential for the state-vector simulator to exploit the data locality in a
single node and minimize the communication across different nodes for efficiency
in practice. What remains challenging, then, is for the simulator to determine
what portion of the computation should be executed locally thus the resulting
amount of communication can be minimized. To achieve this, the quantum circuit
needs to be reorganized systematically.

To this end, we propose \svsim that consists of i)~graph-based quantum circuit
partition approaches that can extensively exploit the hierarchical memory
systems and maintain the efficient communication across nodes, and ii)~a
state-vector simulation framework to support executing the partitioned circuit.
The main workflow of \svsim takes an input quantum circuit that is further
compiled to a directed acyclic graph (DAG), partitions the circuit graph into a
set of smaller circuits (parts), and each smaller circuit is simulated with a smaller
state vector simulator that would fit into the faster memory layer. Contrary to other “partitioning-based” methods~\cite{cutqc,baker2020time} that partition the circuit to execute different parts on multiple real quantum devices (i.e., Quantum Processing Units) in parallel, simulating quantum circuits on classical computers follows a different model. Thus, the computation of each subcircuit (part) occurs on all available resources. The preservation of acyclicity during partitioning and in the resulting quotient graph enables the utilization of the acyclicity property across all components of the simulation framework for better communication and locality patterns.

To the best
of our knowledge, \svsim is among the firsts effort to apply graph-based approaches on the quantum circuit decomposition for efficient circuit
simulation. Unlike other quantum-circuit optimization techniques aiming at reducing the gate counts (i.e., gate fusion), we focus on a global view of the circuit execution and effectively utilize the memory hierarchy. Therefore, our approach is orthogonal and complementary to existing approaches.

Our paper makes the following contributions:

{\tiny$\blacksquare$} We design and implement a novel state-vector simulator: \svsim that
    uses acyclic graph partitioning approaches to take advantage of the multiple
    memory hierarchy in the modern HPC systems to efficiently simulate quantum
    circuits.
    
{\tiny$\blacksquare$} We investigate three graph-based strategies with \svsim's part-based
    execution model and empirically determine the performance benefits offered
    by each strategy.

{\tiny$\blacksquare$} We demonstrate the performance benefits of \svsim on a large-scale
    computing system over the state-of-the-art quantum simulator, and such
    improvement increases with the scaling-up of the number of qubits by up to $\times 3.9$ with $\times 2.1$ on average with \dagp partitioning.

{\tiny$\blacksquare$} We propose a multi-level \svsim model that achieves $\times 1.5$
    improvement over the single-level \svsim 
    (up to $\times 5.7$ over the baseline vs. up to $\times 3.9$ with single-level).

\section{Background and Related work}
\label{sec:bg}
Here, we introduce the basics of quantum computing and how the state vector
simulation performs on the quantum state.

\subsection{Quantum Computing Basics}

A qubit is a block of quantum memory
that consists of any possible quantum superposition of quantum state $\ket{0}$
and $\ket{1}$ as in  $\ket{\phi} = \alpha_{0} \ket{0}+\alpha_{1} \ket{1}$;

where $\alpha_{0}$ and $\alpha_{1}$ are called the quantum amplitudes, such that $\alpha_{0}^{2} + \alpha_{1}^{2} = 1$. Upon measurement, the probability of the whole quantum state collapsing to state 0 is equal to $\alpha_{0}^{2}$, and to state 1 is equal to $\alpha_{1}^{2}$.

A quantum (logic) gate represents an operation on the quantum state, i.e., a
(small) set of qubits. Fundamental quantum mechanics principles state that every
quantum gate is required to be unitary, which means it can be represented
by a $2^{n} \times 2^{n}$ unitary matrix $U$ and $UU^{\dagger} = \mathbb{1}_{n}$, where
$U^{\dagger}$ is the conjugate transpose of $U$ and $\mathbb{1}_{n}$ is the
identity matrix of size $2^{n} \times 2^{n}$. For example, a basic quantum
gate X is equivalent to the classical logical NOT gate. X gate maps the quantum
state $\ket{0}$ to $\ket{1}$ and $\ket{1}$ to $\ket{0}$, and more generally on a
quantum state $\ket{\phi}$:
$X\ket{\phi} = \begin{psmallmatrix}
0 & 1\\
1 & 0
\end{psmallmatrix} \binom{\alpha_{0}}{\alpha_{1}} = \binom{\alpha_{1}}{\alpha_{0}},$
where X =  $\begin{psmallmatrix}0 & 1\\1 & 0\end{psmallmatrix}$. Other common
single-qubit gates include Y~=~$\begin{psmallmatrix}0 & -i\\i &
0\end{psmallmatrix}$, Z~=~$\begin{psmallmatrix}1 & 0\\0 & -1\end{psmallmatrix}$,
H~=~$\frac{1}{\sqrt{2}}\begin{psmallmatrix}1 & 1\\1 & -1\end{psmallmatrix}$, and
multi-qubit gates such as CX, SWAP, etc.

\subsection{State Vector Simulation}

Given an initial quantum state -- a complex-number vector
representing the n-qubit system, and a sequence of quantum gates referred as a
quantum circuit, the state vector simulation conducts the process of applying
each gate inside the quantum circuit on the quantum state, to simulate how the
quantum operations modify the quantum state with the classical computing. The
state vector holds $2^{n}$ elements, occupying a total of $2^{n}\times16 =
2^{n+4}$ bytes (i.e., a complex number requires 16 bytes) of
memory on a classical computer. During the simulation, applying a quantum gate
on the quantum state is equivalent to conducting a matrix multiplication of the
gate matrix on the corresponding positions of the qubit(s) on the state vector
(explained in Sec.~\ref{sec:pattern}).

\subsection{Related Work}

There are several emerging quantum circuit simulation systems focusing on state-vector simulation.
Intel's $q$H$i$PSTER~\cite{qhipster} is a specialized distributed quantum
simulation system that reaches to 42-qubits capacity through various
optimization techniques such as vectorization, multi-threading, cache blocking
and overlapping computation with communication. IQS~\cite{intelqs}, a
recent work of Intel, is the most up-to-date version of $q$H$i$PSTER and one of
the few available open-source simulators. \intel implements the distributed
simulation via careful qubit mapping, to reduce the global communication, yet it
does not present a systematical approach to arrange the local and global qubits,
opposed to what we do in \svsim.

QX~\cite{qx} is a quantum circuit simulation platform which defines its own low-level
language to implement quantum operations.
Similar to $q$H$i$PSTER, it exploits optimization techniques
such as instruction-level parallelism (e.g., SSE, AVX and FMA instructions),
multi-threading, etc. It also leverages
floating point operations reduction and swap-based implementation to conduct the
gate-specific optimizations.

Wu et al.~\cite{wu2019full} employ both lossless and lossy data compression techniques on quantum
simulation to shrink the memory footprint.
During the simulation, certain blocks of data
need to be decompressed for update. It uses the special memory channel residing
in Intel Xeon Phi to efficiently store the decompressed data.

QuEST~\cite{quest} proposes MPI- and OpenMP-based distributed parallel circuit simulation. Alternatively, it leverages the massive parallelism of GPUs to simulate circuits that fit in a single GPU memory. It takes a rather circuit-agnostic approach to
partition the state vector into different MPI ranks, which is close to
\intel's strategy.

Doi et al.~\cite{doi20cache} propose a cache-blocking technique with MPI for Qiskit's state vector simulator Aer. The proposed approaches optimise the quantum circuit itself to improve locality.
H\"aner et al.~\cite{haner-petabyte} reduce the
global communication for the quantum supremacy circuit by swapping the
gates into local state vector simulation. It conducts a closer examination on
the quantum supremacy circuit and identifies the gates that do not need to
communicate between compute nodes.

There are several GPU-based simulators:
NVIDIA recently released cuQuantum SDK~\cite{cuQuantum}
that provides several tools to simulate quantum circuits on GPUs for high-performance circuit
simulation.
SV-Sim~\cite{li21svsim}
is a PGAS-based state vector simulator that simulates a quantum circuit in single/multi- CPU/GPU settings for both single and multi node architectures.
and HyQuas~\cite{zhang2021hyquas}.
The state-of-the-art, HyQuas, partitions the gates in a greedy fashion, which contain no more than a given number of active qubits. HyQuas can switch between different simulation methods for different chunks of a quantum circuit in order to improve performance.

\svsim differs from these work mainly in that it takes a more general and
systematic approach, i.e., the acyclic graph partitioning that determines which qubits and gates can
be gathered for faster simulation. This reduces the communication by allowing each node to execute the subset of the
gates (i.e., gates of current part) on only the local inner state vector.

Moreover, our acyclic graph partitioning technique is orthogonal to
most of
the optimization techniques (e.g., gate fusion, AVX, floating point reduction, etc.)
implemented in above simulators. Hence, our strategies to optimize the
circuit simulation can complete the above techniques and contribute to the
overall simulation efficiency.
Our approach can be used as an encapsulating layer around another off-the-shelf CPU/GPU-based simulator and benefit from their specific optimizations as well (Further details  in Sec~\ref{sec:gpu}).
Suggested by~\cite{cutqc},
smaller-scale quantum computers can be used to execute a portion of the
circuit/qubits in parallel, and our proposed graph partitioning approach may work
for both the actual quantum circuit execution and simulation.

%

\section{\svsim: Hierarchical Quantum Simulation}
\label{sec:hierarchical}

In this section, we
first introduce an analysis on the main computational pattern
of the state-vector (SV) based quantum circuit simulation to lay out the need for the
hierarchical simulation strategy. This motivates the graph-based
circuit partitioning approach that can systematically generate  sub-circuits to
efficiently access the hierarchical memory architecture. Then, we describe our
design and implementation of the hierarchical simulation framework.

\subsection{SV-based Computational Pattern}
\label{sec:pattern}

\begin{figure}
\centering
\begin{subfigure}[b]{0.45\linewidth}
  \centering
  \includegraphics[width=\linewidth]{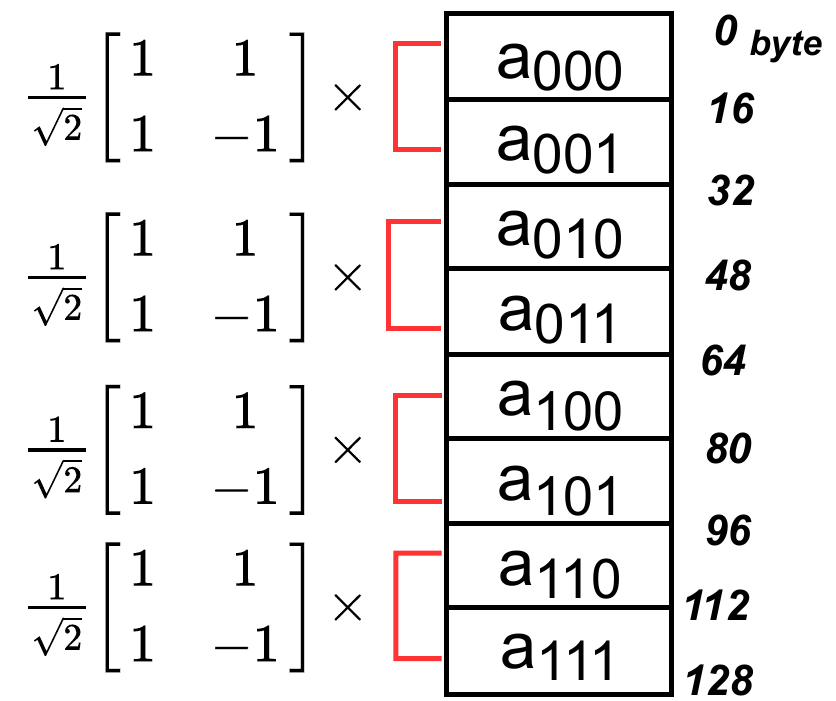}
  \caption{$H$ gate on qubit 0}
  \label{fig:q0}
\end{subfigure}%
\begin{subfigure}[b]{0.48\linewidth}
  \centering
  \includegraphics[width=\linewidth]{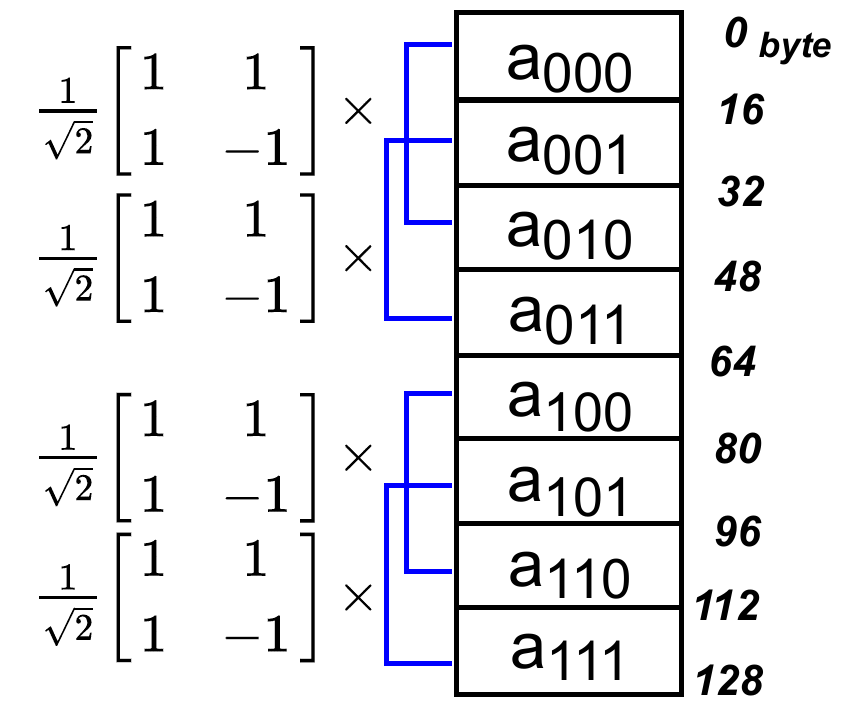}
  \caption{$H$ gate on qubit 1}
  \label{fig:q1}
\end{subfigure}
\caption{Applying an $H$ gate on qubits 0 and 1
of a 3-qubit quantum system ($a_{000}, a_{001}, ... a_{111}$, little-endian).
}
\label{fig:motivation}
\end{figure}

As discussed in Sec.~\ref{sec:bg}, the main computation performed by the
 state-vector simulation is a sequence of ``scoped" matrix multiplications:
 for each gate inside the quantum circuit, the simulator needs to pick up the
 amplitude values stored in the corresponding positions of the quantum state, to
 form a small vector that is multiplied by the gate's matrix. After the
 multiplication, the elements of the resulting vector are stored back to where
 they are extracted from in the state. Fig.~\ref{fig:motivation} illustrates
 the process of applying $H$ gate on qubits~0 (Fig.~\ref{fig:q0}) and~1 (Fig.~\ref{fig:q1}) in a 3-qubit quantum system.
 In both examples, the total simulation of the gate involves 4 independent matrix-vector multiplications.
 Each matrix-vector multiplication involves the $2 \times 2$ $H$-gate matrix and a two-element vector extracted from the 8-element state vector, where each element is 16 bytes in size.
In general, for applying a single
qubit gate such as $H$, the total number of matrix multiplications needed is
$2^{(n-1)}$. One matrix multiplication requires 4 complex $\times$ operations
and 2 complex $+$ operations. In addition, multiplying two complex numbers takes
4 \texttt{FP} $\times$ and 2 \texttt{FP} $+$, thus the total number of FLOPS are
28 for one matrix multiplication. Since the data transferred between DRAM and
cache is 64 bytes ($16\!\cdot\!2\!\cdot\!2$) per one matrix multiplication, the operational
intensity~\cite{roofline} is $\frac{7}{16}$, indicating that the entire
computation is constrained more by the data movement (i.e., memory bound)~\cite{roofline}.

Efficient quantum circuit simulation requires optimizing data locality
in performing the matrix multiplication operations. One observation on the
computational pattern suggests that this task follows a relatively
cache-friendly pattern: a pattern shown in Fig.~\ref{fig:q0} conducts
sequential memory accesses (i.e., step size $s=1$); the pattern shown in
Fig.~\ref{fig:q1} (i.e., $s=2$) would exhibit a similar cache efficiency as the
Fig.~\ref{fig:q0} since a step size of 2 would allow the access to the
vector's elements to remain in the same cache line (typically 64 bytes for modern
CPUs). In fact, the step size $s$ for picking up the two elements of the vector
is determined by the position $i$ of the target qubit\footnote{It
is the case for the single-qubit gate. For multi-qubit gates such as
multi-control gates, the implementation relies on the gate decomposition to
convert it to the single-qubit case with a proper offset.}, i.e., $q_{i}$, accessed by the
quantum gate can be computed as $s=2^i$.

However, as the number of qubits needed by the quantum circuit scales up,
the working set size of the simulation task would inevitably exceed the cache size. For a modern CPU usually featuring a L3 cache (e.g., 32MB) shared by all
cores, a L2 cache (e.g., 1MB) per core and a L1 cache (e.g.. 64KB) per core, the
performance of the circuit simulation on more than 21 qubits (i.e.,
$2^{21}\!\cdot\!16=32$MB) would be degraded by the capacity cache misses on L3 cache. 

\subsection{Hierarchical Circuit Simulation}
To address the above challenge and improve the performance of 
quantum circuit simulation, \svsim implements the hierarchical simulation framework that executes the input circuit as a
sequence of sub-circuits (i.e. parts), each of which contains a portion of the original gates.
Therefore, executing the gates inside each part would only occupy a smaller
working set size in the memory, which presents a better locality compared to the
non-hierarchical scenario.

\svsim makes the following design choices: a) \svsim considers the dynamic execution of the gates in a
quantum circuit as a DAG, and converts the circuit
partitioning problem into a (acyclic) graph partitioning problem (Section~\ref{sec:partitioning}); b) \svsim ensures the correctness of the proposed part-based simulation via the novel \textit{Gather-Execute-Scatter} model (Section~\ref{sec:single}); c) \svsim enforces a consistent layout for qubit organization across multiple levels in the hierarchy and remaps the qubits in a part to the lower level simulation.

\subsection{\svsim Framework Overview}
\label{sec:single}
\begin{figure*}[th]
\centering
\begin{subfigure}[b]{0.99\textwidth}
\centering
  \includegraphics[width=0.9\textwidth]{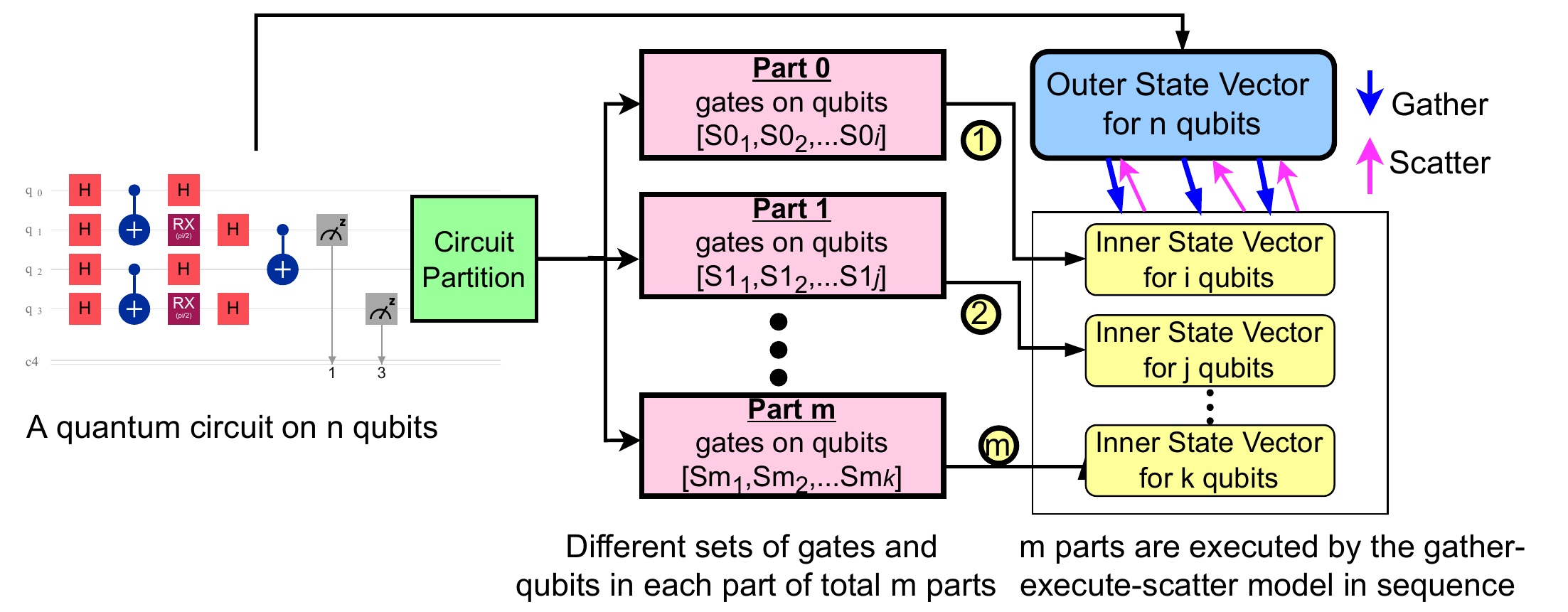}
  \caption{\svsim workflow overview. The input circuit is converted
    into multiple parts via the partition module, and the resultant
    parts of the original circuit to create smaller inner state vector instances. For each of them, the gather-execute-scatter model is applied against the outer state vector that is created for the input circuit.}
\label{fig:overview}
\end{subfigure}
\begin{subfigure}[b]{0.99\textwidth}
\centering
  \includegraphics[width=0.5\textwidth]{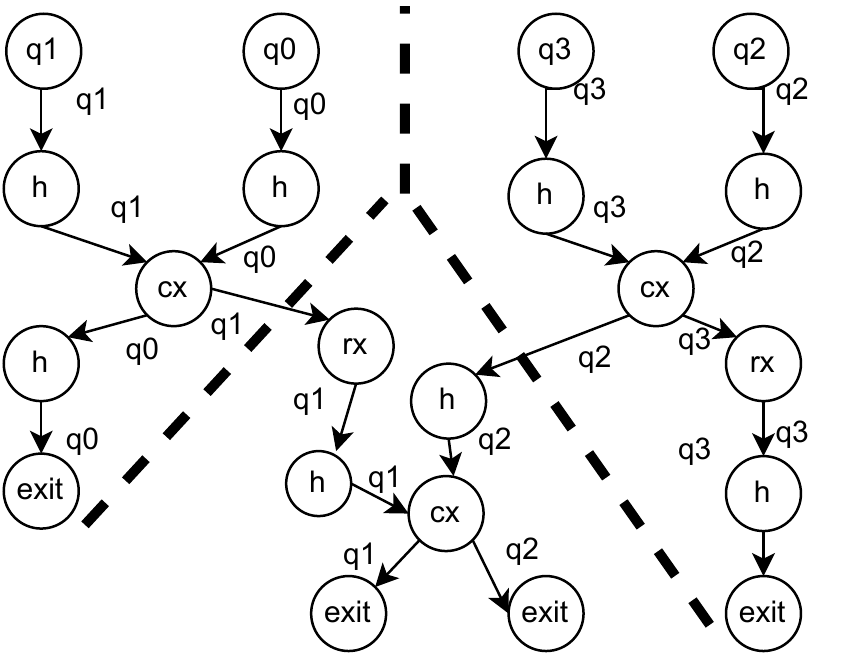}
  \caption{The DAG representation of the example circuit shown in
    Fig.~\ref{fig:overview}. The dotted lines illustrate the desirable partitions of the
    graph.}
\label{fig:example}
\end{subfigure}
\caption{The main \svsim workflow to partition the input circuit
    and perform part-based quantum circuit simulation.}
\label{fig:main}
\end{figure*}

Fig.~\ref{fig:overview} illustrates the overview of \svsim framework. Given a
quantum circuit that is comprised of a sequence of gates applied to $n$ qubits,
\svsim uses the circuit partition module to parse the circuit and generate
several parts/subcircuits of the circuit. Each of the resulting parts of the
circuit would impact a smaller number of qubits than the original circuit,
enabling the hierarchical simulation that the new instances of the state-vector
simulator can be launched over the parts of the circuit. During the simulation,
the amplitudes in the corresponding positions of the states with greater strides
(i.e., handled by the ``outer'' state vector simulator) are gathered into
the low-level quantum states (i.e., handled by the ``inner'' state vector
simulators), and after the execution of the gates inside that part, the result
amplitudes are scattered back to their original positions in the outer
state vector for the next round of gather-execute-scatter operations.

For example, to partition the circuit shown in Fig.~\ref{fig:overview}
(Fig.~\ref{fig:example} depicts the circuit's DAG), one partition
strategy is to divide the circuit into three parts: part 0 (left), part 1 (right) and part 2 (bottom-middle), where each part contains 2 qubits in the working set. 
This way, both the gate dependencies
and the qubit dependencies
are preserved.
While the simulation still sweeps through the entire ``outer'' state vector, the actual operation occurs on 4-element state vectors. In general, such cache-resident inner state-vectors can reduce accesses to data that resides in DRAM from the outer state vector, reducing overall memory access costs.

The workload distribution for the algorithm is as follows:
First, the DAG is generated and partitioned.
Second, the partitioning information is used to decide the computation of parts sequentially (i.e.,
workload in different parts are not computed in separate processing elements).
For a part at hand, a subset of the elements in the original state vector (i.e. outer state vector) is designated, and
\emph{gathered} to ``inner state vector''s. These are disjoint and collectively exhaustive over the
whole state vector.
Within the part, all gates belonging to that part are computed against the inner state vector.
After the computation of a part, the results are \emph{scattered} back to the ``outer state vector''.
And the distinct state vectors for the next part is loaded into the inner state vector for computation.

Algorithm~\ref{alg:gather_scatter} describes how the quantum states get
gathered and scattered between the outer and inner state vectors, using
the example circuit shown in Fig.~\ref{fig:overview}.
To execute the first
part (i.e., $p = P_0$), an inner state vector \texttt{in\_sv} is created based
on the qubits residing in $P_0$ (i.e., $[S0_{1},S0_{2},...S0_{i}]$ as shown in
Fig.~\ref{fig:overview}). Given $S0_{1} = q1$ and $S0_{2} = q0$
, the qubits
that do not participate in any gates in $P_0$ are $q2$ and $q3$. For all combinations of bit
values of $q2$ and $q3$, 
\svsim moves the data addressed by
$[q3,q2,q1,q0]$ from \texttt{out\_sv} to the corresponding $[q1,q0]$ positions
in \texttt{in\_sv}.
This process is referred as \textit{Gather}.
The \textit{Scatter} process sends the states back to \texttt{out\_sv}
after the execution of the gates through \texttt{in\_sv}.
On completion of a number of $2^{t}$ Gather-Execute-Scatter processes (assuming t equals to the number of qubits not residing in the current part),
\texttt{out\_sv} is ready for the next part.   

\begin{algorithm2e}
\DontPrintSemicolon
\SetKwInOut{Input}{Input}
\begin{small}
\Input{
out\_sv = state\_vector(num\_count),\\
Partitions = an acyclic partitioning of gates
}
\For{$p$ in Partitions}{
Get \texttt{gate\_list} from $p$\;
w = working set size \tcp{inner qubit count}
in\_sv = state\_vector(w) \tcp{initialize with size w}
\tcp{e.g., binary permutations of 2 qubits = \{(0,0),(0,1),(1,0),(1,1)\}}
\For{each binary permutation of ($q_{n-1}$, \dots, $q_w$)}{
    \tcp{Gather}
    \For{each binary permutation of ($Sp_1$, \dots, $Sp_w$) }{
        in\_sv[$Sp_1$, \dots, $Sp_w$] \textleftarrow   out\_sv[$q_{n-1}$, \dots, $q_w$, $Sp_1$, \dots, $Sp_w$]\;
    }
    executeSimulation(in\_sv, \texttt{gate\_list})\;
    \tcp{Scatter}
     \For{each binary permutation of ($Sp_1$, $Sp_w$) }{
        out\_sv[$q_{n-1}$, \dots, $q_w$, $Sp_1$, \dots, $Sp_w$] \textleftarrow  in\_sv[$Sp_1$, \dots, $Sp_w$]\;
    }
}
}
\vspace{-0.6em}
\caption{The algorithm to execute a circuit given an acyclic partitioning.}
\label{alg:gather_scatter}
\end{small}
\end{algorithm2e}



\subsection{Multi-node Design}
\label{sec:multi}

For circuits that require a large number of qubits,
a single compute node will not be able to hold the entire quantum state vector in memory.
\svsim extends our hierarchical simulation design to develop a multi-node MPI-based hierarchical simulation for distributed-memory quantum circuit simulations.
The first task to achieve this is to distribute the quantum states from the state vector across all MPI ranks to ensure that each MPI rank can execute all the gates inside a part locally.
Since different parts may contain different qubits, to switch between parts,
each MPI rank needs to gather the states residing in the remote MPI ranks to form new local states,
which triggers a global communication across MPI ranks.

\begin{figure*}[t]
     \centering
     \begin{subfigure}[t]{0.48\textwidth}
         \includegraphics[width=\textwidth]{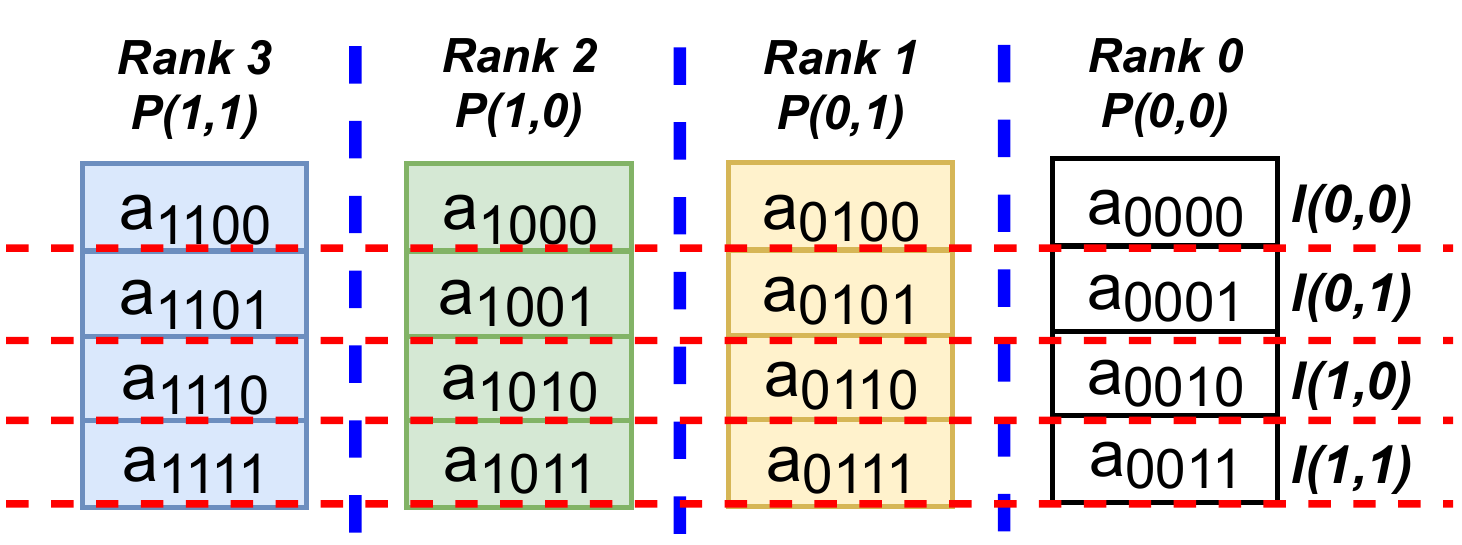}
        \caption{The initial state distribution across 4 MPI ranks.}
        \label{fig:mpi_init}
     \end{subfigure}
     \hfill
    \begin{subfigure}[t]{0.48\textwidth}
         \includegraphics[width=\textwidth]{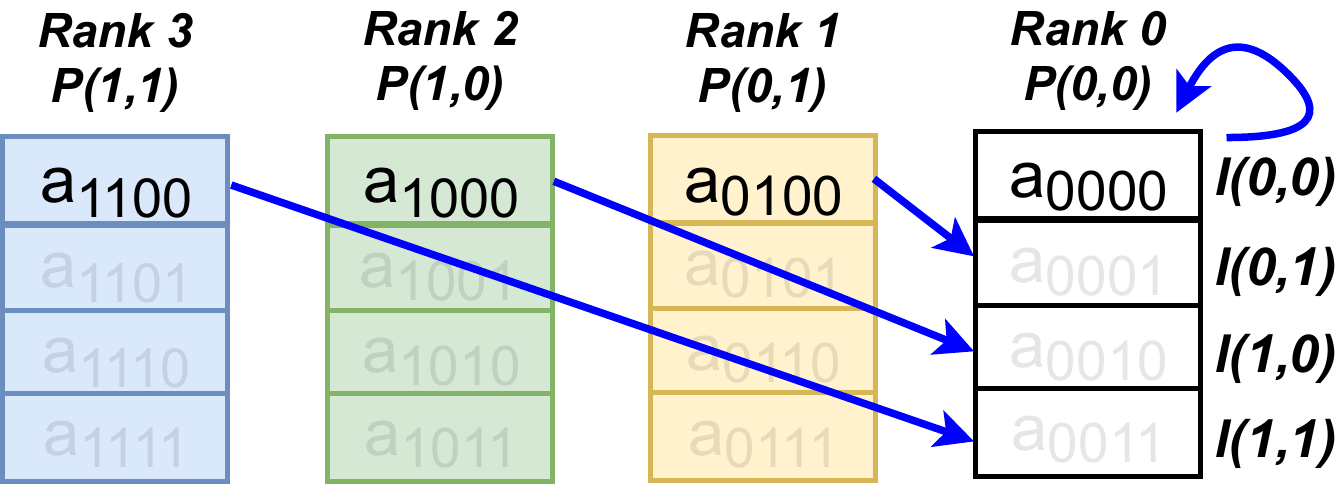}
         \caption{Example: the data gathered by rank 0 switching from \texttt{Part 0} to \texttt{Part 1}.}
         \label{fig:mpi_comm0}
     \end{subfigure}
          \hfill
     \begin{subfigure}[t]{0.48\textwidth}
         \includegraphics[width=\textwidth]{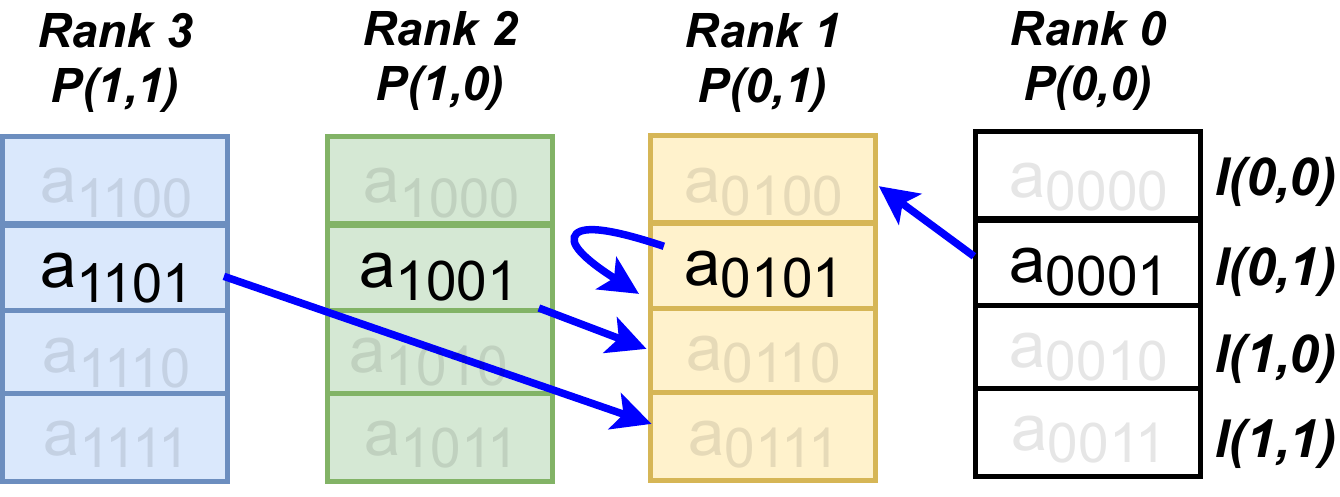}
         \caption{Example: the data gathered rank 1 switching from \texttt{Part 0} to \texttt{Part 1}.}
         \label{fig:mpi_comm1}
     \end{subfigure}
          \hfill
     \begin{subfigure}[t]{0.48\textwidth}
         \includegraphics[width=\textwidth]{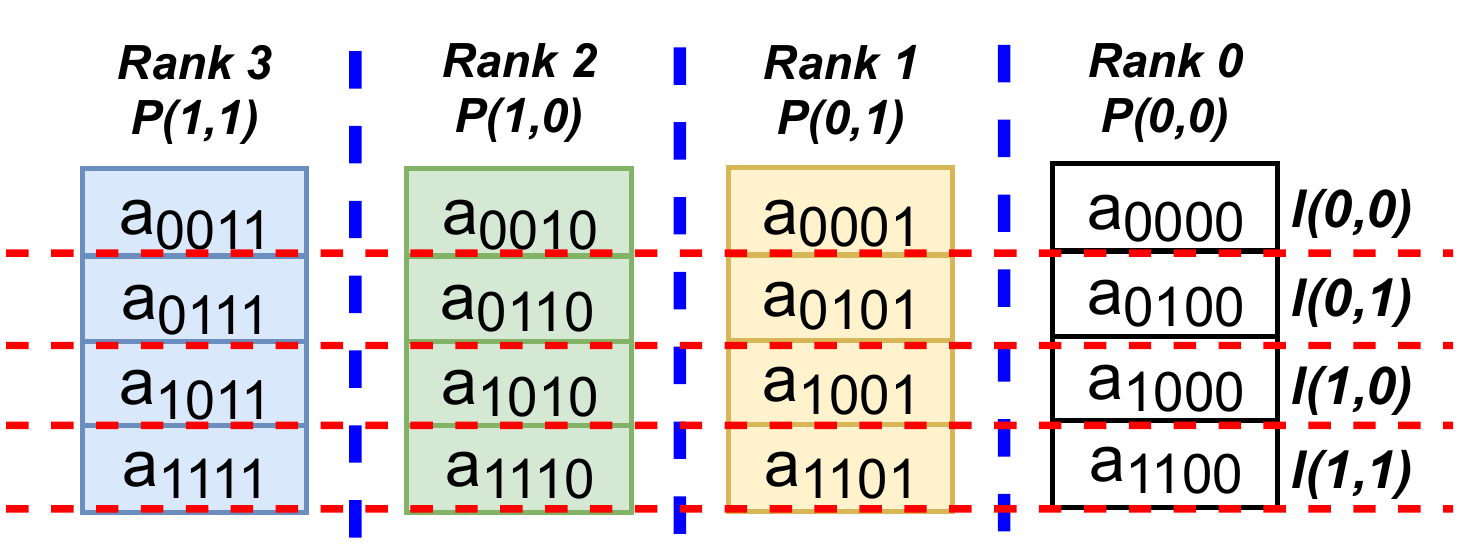}
         \caption{Final distribution of the state vector amplitudes after across-rank communication.}
         \label{fig:mpi_full}
     \end{subfigure}
     \caption{The examples of MPI messages across MPI ranks to gather and update local state vector of rank 0 and 1. }
     \label{fig:mpi}
\end{figure*}

Specifically, assuming an $n$-qubit quantum system, \svsim identifies the qubits as two sets:
the number of $p$ process (or MPI rank) qubits and the number of $l$ local qubits,
where $n = p + l$.
Note that this design requires the number of MPI ranks to be a power of two\footnote{This constraint can be relaxed with virtual ranks and mapping multiple virtual ranks to MPI ranks, we do not address this issue in this work.}.
For example, a 4-qubit quantum state vector simulated by 4 MPI ranks (i.e., $p=2$ and $l=2$) would be addressed as $[a_{3}, a_{2}, | \ a_{1}, a_{0} ]$.
The 4 MPI ranks are addressed by the most significant bits $(a_{3}, a_{2})$,
with the local quantum state elements of each MPI rank addressed by $(a_{1},a_{0})$. Fig.~\ref{fig:mpi_init} illustrates the MPI-based layout of the quantum states for the scenario when part $P_0$ of the example circuit
(shown in Fig.~\ref{fig:overview}) is executed by \svsim.
After $P_0$ has been executed, \svsim uses the new local and process qubits designated for the next part
to determine the state to be gathered for each MPI rank.
The examples shown in Fig.~\ref{fig:mpi_comm0} and Fig.~\ref{fig:mpi_comm1} illustrate the MPI messages received by rank 0 and rank 1, respectively,
and the amplitude values and locations in the send/receive buffers after $P_0$.
Fig.~\ref{fig:mpi_full} shows the final amplitude distribution across all MPI ranks for the new part ($P_1$) to be executed,
where the process qubits are $(a_{1},a_{0})$ and the local qubits are $(a_{3},a_{2})$.

Our MPI-based implementation to distribute and gather qubits in the multi-node scenarios offers a general interface for other simulators to use as a library.

\section{Quantum Circuit Partitioning}
\label{sec:partitioning}
The memory requirement for a given number of qubits is exponential and each
iteration of the simulation starts and ends with memory read/writes between the inner and outer state vectors
(Alg.~\ref{alg:gather_scatter}).
Thus, an intuitive partitioning approach would be to minimize the number of
parts, i.e.,  minimizing the number of bulk
read/writes (or MPI communications) of this exponential amount of data. 
Hence, the objective of the partitioning is to create the smallest
number of parts while keeping the number of qubits
of each part under a limit. If there are
cyclic dependencies between parts (i.e., parts require computation of gates from
each other), then it would require multiple data transfers to synchronize. Therefore, having cyclic dependencies makes it
hard to count and minimize the number of bulk read/writes and MPI communications. Thus, a circuit should be
partitioned in a way that allows each part to be loaded into the inner state
vectors for computation only once after all of its dependencies
are met, i.e., an acyclic manner with a topological order.
And,  finding such a
partition should be fast and efficient to potentially bring runtime improvement.

\subsection{Model}
\label{sub:partitioning.model}
We consider a directed acyclic graph (DAG)  $G=(V,E)$,
where the vertices in set~$V=\{v_1, \ldots, v_n\}$
represent computational gates,
and edges represent the qubits needed for the computational gates.
Given $v_i\in V$, $pred_i = \{v_j \; | \; (v_j, v_i)\in E\}$
is the set of predecessors of gate~$v_i$ in the graph,
and  $succ_i = \{v_j \; | \; (v_i, v_j)\in E\}$ is the set of successors of gate~$v_i$.
We create artificial computational gates for initialization and destruction of the qubits,
i.e., for each qubit, there is an {\em entry} and an {\em exit} gate that does not represent any computation.
{\em Entry} gates have no predecessor and one successor
that is the first computational gate the corresponding qubit enters.
And, {\em exit} gates have no successors and one predecessor.

For each computational gate, the total incoming edge weight is equal to the outgoing edge weight, which is the number of qubits involved in the computation of this gate.
No qubit is input to multiple gates at the same time.
Thus, it is possible to trace each qubit through the edges between gate vertices.
The gates are naturally in a topological execution order where they can be executed only after all of their predecessors are executed,
i.e., the edges represent the \emph{qubit dependency} between the gates.

The {\em Circuit Partition} (Fig.~\ref{fig:overview}) accepts a DAG representation of the circuit, and a maximum working set size~$L_m$ that is common for all parts as input.
It solves the acyclic $k$-way partitioning problem $P=\{V_1, \ldots, V_k \}$ of the DAG $G=(V,E)$: the
set of vertices~$V$ is divided into $k$ nonempty, pairwise disjoint, and collectively exhaustive {\em parts} satisfying three conditions:
i)~The working set size of individual parts do not exceed $L_m$.
ii)~The partition is acyclic.
There is a path between $V_i$ and $V_j$
($V_i \leadsto V_j$)
if and only if there is a path between a vertex~$v_i \in V_i$ and a vertex~$v_j\in V_j$.
The {\em acyclic} condition means that given any two parts $V_i$ and $V_j$, we cannot have
$V_i \leadsto V_j$ $and$ $V_j \leadsto V_i$.
A \emph{part-graph} of $G$ is a graph where the nodes represent the parts in $G$ (i.e., all nodes in the same part are contracted to a single node) and the edges represent the cumulative edges between parts of $G$.
iii)~The number of parts $k$ is minimized.
Similar to the graph partitioning problem, the acyclic DAG partitioning is an NP-Hard problem and there is no k-approximation for $k > 2$~\cite{Herrmann19-SISC}. Thus, all feasible algorithms that solve this problem are heuristics.
The output of partitioning is a part assignment for each node (gate). After partitioning, gates are collected
according to their parts and executed with respect to the original order among
those in the same part.


The \emph{working set size}, $L(V_i)$ of a part $i$%
, is the number of qubits needed
for all gates in that part. That is, if gate $A$ needs $q0$ and $q1$ and gate $B$
needs $q0$ and $q2$, these two gates require 3 qubits in total. If $V_i =
\{A, B\}$, then $L(V_i)$ = 3. We can imagine edges having labels in the
computational gate graph. Each entry qubit emits an edge with the label of qubit
they are initializing, and each gate these qubits participate in, they are
represented by a single in-edge and a single out-edge. Since a qubit cannot be
passed to two different gates at the same time (i.e., gates have a time order),
the incoming edges of a gate/part are always unique.

\subsection{Proposed Partitioning Methods}
We propose three partitioning approaches:

\subsubsection{Natural Topological Order Cutoff (\nat)}
\nat follows the execution order of the gates in the original circuit,
and computes the working set size.
When the working set size exceeds the limit $L_m$,
the gates before the limit is exceeded are assigned as one part.
Then, working set size is reset and this process is repeated for the remaining gates.
Natural ordering is deterministic and falls short when the order contains alternating operations for larger number of qubits than $L_m$.

\subsubsection{DFS Topological Order Cutoff (\dfs)}
\dfs remedies this problem
by testing several random DFS topological orders of the gates
instead of the deterministic natural topological order,
and picks the one that yields the smallest number of parts.
The part assignment follows the same procedure.

\subsubsection{Acylic-partitioning-based Partitioning (\dagp)}
We propose a novel acyclic DAG-partitioning-based heuristic
by extending and modifying a recent open-source, state-of-the-art acyclic DAG
partitioner~\cite{Herrmann19-SISC,Ozkaya19-PPAM}.
It presents an acyclic DAG partitioning
algorithm that consists of an acyclic agglomerative clustering,
recursive-bisection-based initial partitioning, and refinement phases. Given a
DAG and $k$, their algorithm recursively divides the DAG, until the desired $k$ is achieved,
while minimizing the edge cut (total weight of edges that connect nodes from different parts) as the objective.
Using a black-box partitioner for our problem definition is not trivial. There are major changes to
the partitioning objective, criteria, and greedy approaches as well as some minor changes in order to preserve/improve
the performance of a partitioner. Without these modifications, the partitioners may not ever find a
valid partitioning abiding the required constraints.
We replaced the edge cut objective with minimizing the number of parts,
while ensuring each part's working set size is under given limit.
We modified related computations in all phases of the partitioning.
We added a final merging phase that is not present in the original algorithm.
We used the authors' suggested parameter values except for the imbalance ratio (i.e., $\epsilon \leq 1.5$) since
the weight balance of the parts is not critical. We refer the readers to the
corresponding article for further details on the original algorithm and only briefly
discuss our major modifications.

The original algorithm requires the final number of parts, $k$, as an input parameter.
Our problem requires the algorithm to discover the necessary (minimum) number of parts.
Our algorithm starts by computing the working set size of the graph $L(G)$
(it is the total number of qubits, i.e., number of $entry$ nodes).
If it is greater than $L_m$, the graph is partitioned into two roughly balanced subgraphs (in terms of the number of nodes).
The recursive bisection operation dives into the subgraphs and partitions them until each one's working set size is less than or equal to $L_m$.
If the working set size of a subgraph is less than or equal to $L_m$, then the partitioner
stops and returns.
At the end of recursive bisection, we run
a merge phase in which a clustering algorithm is applied on the global view of the graph via the \emph{part-graph}
to merge existing parts until there is no more possible valid mergers,
i.e., the merger does not create cycles in the
\emph{part-graph} and does not violate the $L \leq L_m$ criterion. 

The runtime performance of the partitioner significantly depends on how efficiently
one can compute updates when a vertex is moved from one part to another. 
The edge cut, objective of the former problem~\cite{Herrmann19-SISC,Ozkaya19-IPDPS}, can be efficiently updated.
Our \dagp algorithm takes advantage of the following information to compute working set size at each phase efficiently.
Since the computational gate graph of quantum circuit simulation has the property
where the number of inputs of a gate is equal to that of outputs  and all of them represent unique qubits,
one only needs to count the in-edges to a part, and the number of $entry$ nodes within the part.

One question that may arise is whether the circuits with higher number of entanglements with many
qubits complicates the partitioning. The partitioner considers the circuits as DAGs
where the edges contain all dependency information
and \svsim makes no assumptions on the properties of the circuits. Larger entanglements are just
nodes with higher degrees. This may deteriorate the performance for the greedy
localized approaches such as \nat and \dfs for finding the best valid partitions. The goal of \dagP
approach is to find the best partitioning using a global view of the computation DAG, which
overcomes this deficiency of the former two approaches.

Both \nat and \dfs methods have a localized view of the gates at hand whereas
\dagp benefits from a broader view of the whole circuit graph.
Compared to the runtime of the quantum circuits, all three have negligible computation times (up to milliseconds).

\begin{figure*}
    \centering
    \includegraphics[width=0.33\textwidth,height=25em]{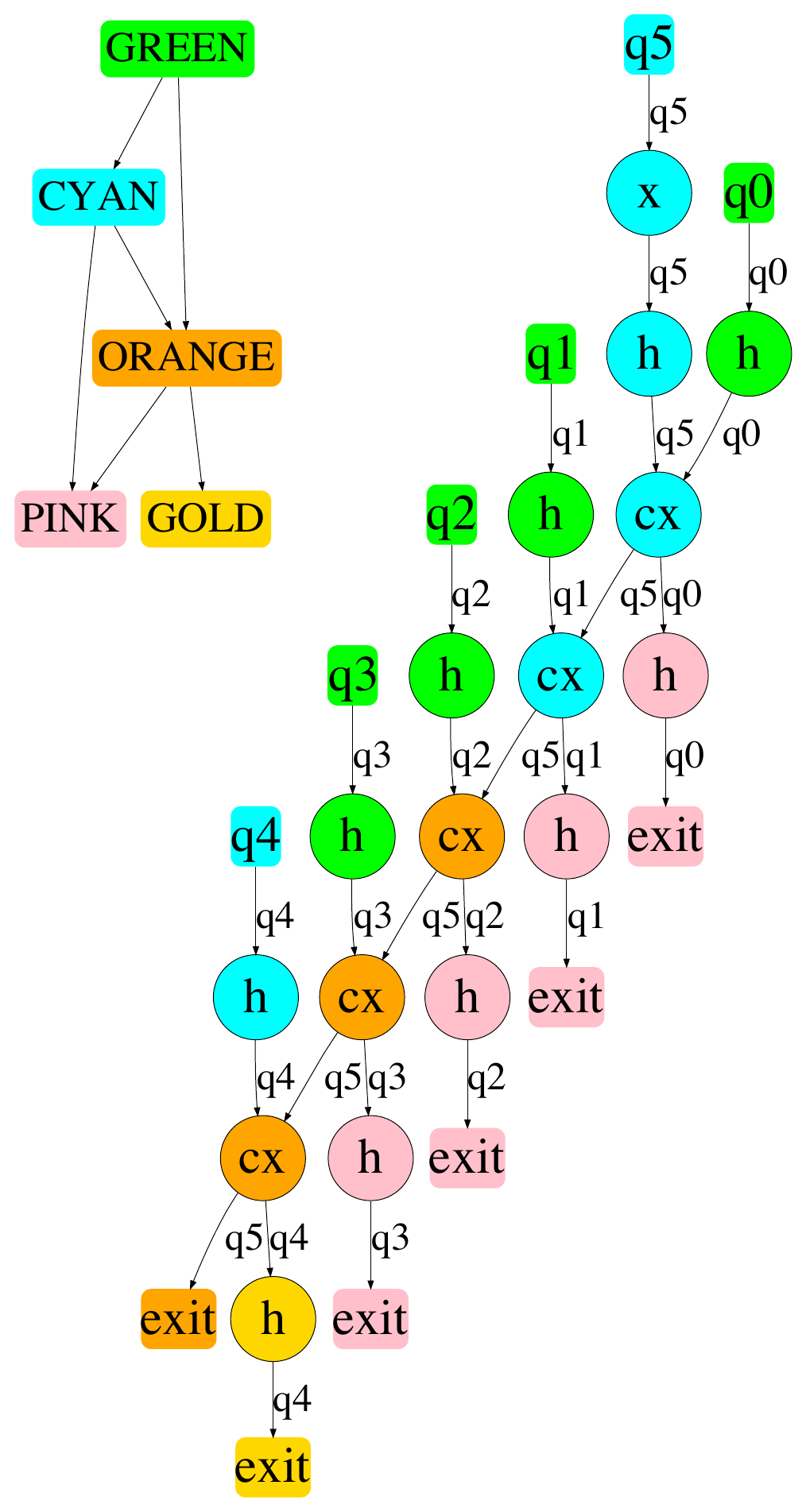}
    \includegraphics[width=0.36\textwidth,height=25em]{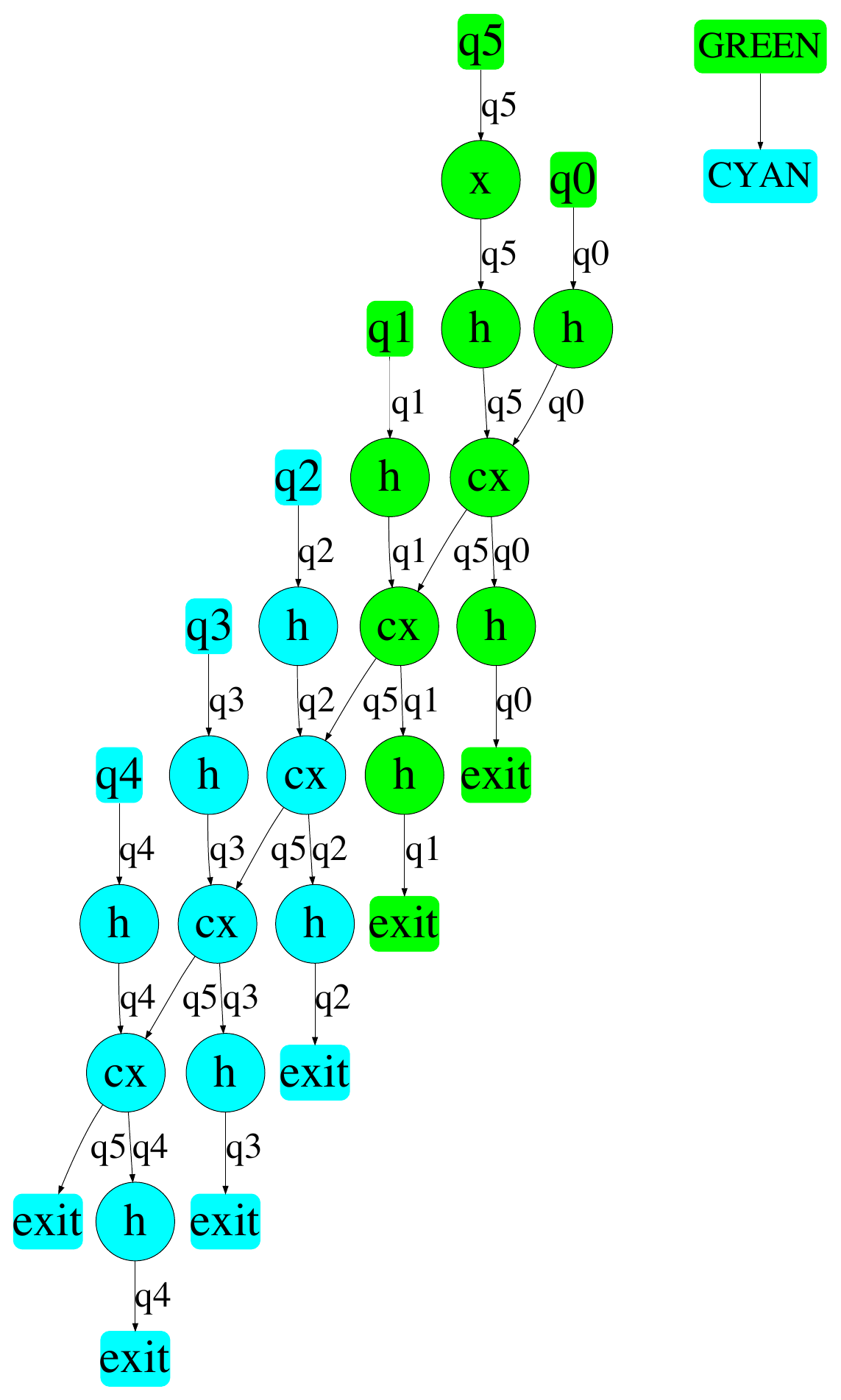}
    \caption{
    A toy example partititoning of bv graph with 6 qubits with qubit limit 4 using \nat (left) and \dagp approach (right).
    }
    \label{fig:partitioning_nat_dfs}
\end{figure*}

Figure~\ref{fig:partitioning_nat_dfs} shows a toy example for partitioning with \nat and \dagP approaches, respectively.
Top-left and top-right subgraphs show the respective \emph{part-graphs}.
\dfs approach can return any number of parts between these two examples depending on the random DFS topological orders at hand.


\textbf{Multi-level partitioning.}
The acyclic DAG partitioning with recursive bisection presents an opportunity to
prepare partitions for varying $L_m$'s at different scales, and thus, can be
better exploited for the multi-node design (Sec.~\ref{sec:multi}), i.e.,
inter-node data distribution and intra-node cache locality optimization.
The partitioning is run on the original circuit graph with the first-level limit
to break it down into a number of parts, then, each one is recursively
partitioned further using the second-level limit.

The $L_m$'s are decided with the system configuration in mind.
The first-level partitioning uses a $L_m = l$ for the local state vector (Sec.~\ref{sec:multi}),
while the second-level partitioning uses a size that keeps the size of the second-level inner state vector under the LLC cache size.
When the number of qubits in a part is less than $L_m$,
we add the qubits from the higher level part to exploit spatial locality.
\section{Experimental Evaluation}
\label{sec:method}
Table~\ref{tab:benchmark} lists the 13 benchmark quantum circuits, from QASMBench suite~\cite{qasmbench}, used in this paper. Circuits cover a variety of application domains that represent the essential areas in quantum computing.
Note that we explore two sets of qubit and gate configurations for the bv, cc, and ising circuits to investigate how much more efficiency our \svsim could bring to the same quantum algorithms in different scales.

We ran our experiments on a workstation (Intel Cascade Lake, 448 CPU cores, 8 sockets, 8 NUMA nodes, 6TB memory)
and the Frontera~\cite{frontera} supercomputing cluster at TACC (each node in the cluster features dual Intel Xeon Platinum 8280 with 2.7GHz clock rate, 56 cores, 192GB DRAM and the InfiniBand HDR-100 network).
For each circuit, we examine the performance of the simulation with three partitioning strategies (\nat, \dfs, and \dagP) 
and compare the results with the baseline simulation results using Intel's \intel~\cite{intelqs}.
We translate the OpenQASM gates into \intel gates to ensure that \svsim and \intel execute identical circuits.

\subsection{Single node experiments}
On the single machine scenario, we configure the \svsim with varying number of OpenMP threads (i.e., 2,4,8,16,32,64,128),
and \svsim exhibits a close-to-linear speedup in this strong scaling case.
Table~\ref{tab:cache} shows the breakdown of the memory usage (profiled with \textit{vTune Profiler}~\cite{vtune})
with a single thread for each partitioning strategy to indicate why \dagP offers the most efficient simulation.
Among three strategies, \dagP always leads to the lowest DRAM stalled time, which reflects the more efficient cache access pattern.
For the example circuits presented in Table~\ref{tab:cache}, 35.7\% and 20.2\% of pipeline slots
are for memory loads and stores in \nat partitioning on bv and ising respectively, which far outweighs the \dfs and \dagp.
In particular, \dfs and \dagp on ising produce similarly partitioned circuits, hence the memory access patterns are comparable.
Similar trends can be observed in all circuits but qpe (where \nat outperforms \dfs and \dagp).
For brevity, we omit the rest of the results.

To evaluate the quality of the \dagP heuristic,
we implemented an integer linear programming (ILP)-based optimal solution to our modified acyclic DAG partitioning problem.
The ILP solution takes minutes for even the smaller circuits compared to the microseconds of the \dagp heuristic
since the problem is NP-Hard and ILP formulation gets exponentially more complex.
Out of 52 combinations (13 inputs, 4 qubit limits),
\dagP finds the \emph{optimal} number of parts for 48 cases and only differs by 1 or 2 for the rest.


\begin{table}[t]
\centering
\caption{Benchmark description.}
\begin{tabular}{@{}ll@{}rrr@{}}
\toprule
\textbf{Circuit} & \textbf{Description} & \textbf{qubits} & \textbf{gates} &\textbf{Mem.}\\ \toprule
cat\_state~\cite{cat}~ & Coherent superposition  & 30 & 60 & 16 GB\\
bv~\cite{bv} & Bernstein-Vazirani algorithm & 30 & 102 & 16 GB\\
qaoa~\cite{qaoa} & Quantum approx. optimization & 30 & 1,380 & 16 GB\\
cc~\cite{cc} & Counterfeit coin finding & 30 & 149  & 16 GB\\
ising~\cite{ising} & Quantum simulation for ising model & 30 & 354 & 16 GB \\
qft~\cite{qasmbench} & Quantum Fourier transform & 30  & 2,235 & 16 GB\\
qnn~\cite{qasmbench} & Quantum neural network & 31 & 164 & 32 GB\\
grover~\cite{grover} & Grover’s algorithm & 31 & 207 & 32 GB \\
qpe~\cite{qpe} & Quantum phase estimation & 31 & 5,731 & 32 GB\\ \midrule
bv35~\cite{bv} & Bernstein-Vazirani algorithm & 35 & 119 & 512 GB\\
ising35~\cite{ising} & Quantum simulation for ising model & 35 & 414 & 512 GB \\
cc36~\cite{cc} & Counterfeit coin finding & 36 & 106 & 1 TB\\
adder37~\cite{adder} & Quantum Ripple-Carry adder   & 37 & 154 & 2 TB\\
\bottomrule
\end{tabular}
\label{tab:benchmark}
\end{table}

\begin{table}[]
\centering
\caption{Memory access breakdown.}
\begin{tabular}{@{}llrrrrr@{}r@{}}
\toprule
\multirow{2}{*}{\textbf{Circuit}} & \multirow{2}{*}{\textbf{Strategy}} & \multicolumn{4}{c}{\textbf{\% of clockticks}}           & \multirow{2}{*}{\textbf{\begin{tabular}[c]{r}(\%)Memory/ \\ Pipeline slots \end{tabular}}} & \multirow{2}{*}{\textbf{\begin{tabular}[c]{r}Execution \\ time (s)\end{tabular}}} \\
 & & \textbf{L1} &\textbf{L2} & \textbf{L3} & \textbf{DRAM} & & \\
\toprule
& Nat & 6.1  & 4.0 & 4.4 & 19.8 & 35.7 & 209.7 \\
bv & DFS & 2.3 & 3.1 & 3.8 & 16.6 & 26.1 & 172.8 \\
& dagP & {2.9} & {6.5} & {2.0} & \textbf{4.3} & \textbf{20.9} & \textbf{163.2} \\\midrule
& Nat & 7.0  & 2.7 & 4.4 & 11.2 & 20.2 & 613.5 \\
ising & DFS & 1.5 & 1.2 & {1.9} & 5.8 & \textbf{6.6} & 455.6 \\
& dagP & {1.3} & {1.2} & 2.1 & \textbf{5.5} & 7.5 & \textbf{454.1} \\
                                 \bottomrule         
\end{tabular}
\label{tab:cache}
\end{table}

\vspace{-1em}
\subsection{Multiple node experiments against \intel}
\label{sec:eval}

\begin{figure*}
     \begin{subfigure}{\textwidth}
     \centering
         \includegraphics[width=0.61\textwidth]
         {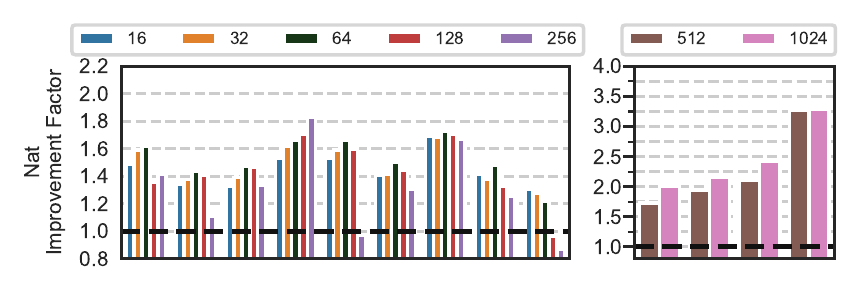}
        \label{fig:Nat_improv}
    \vspace{-2ex}
     \end{subfigure}
     \begin{subfigure}{\textwidth}
          \centering
         \includegraphics[width=0.6\textwidth]
         {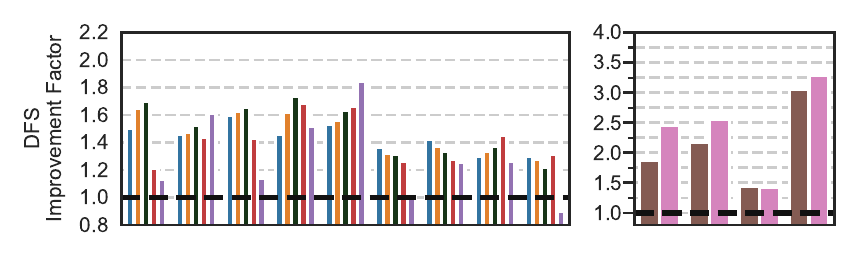}
        \label{fig:DFS_improv}
    \vspace{-2ex}
     \end{subfigure}
     \begin{subfigure}{\textwidth}
          \centering
         \includegraphics[width=0.61\textwidth]{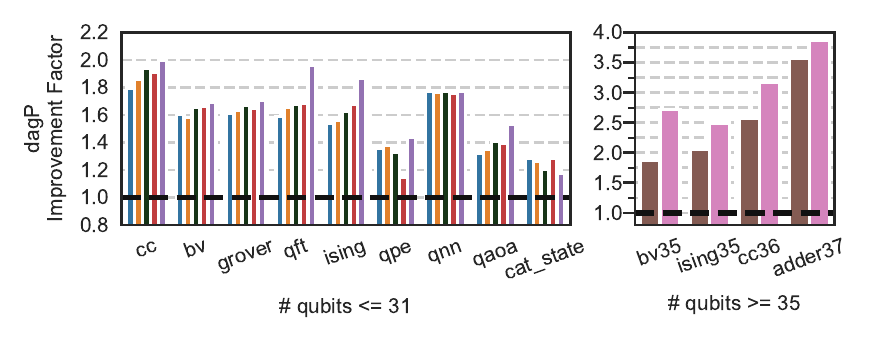}
        \label{fig:dagP_improv}
     \end{subfigure}
     \caption{The improvement factor over Intel IQS of our partition strategies for different MPI ranks.}
     \label{fig:improve}
\end{figure*}

In this section, 
we increase the number of nodes from 16 to 256, on Frontera system.
The circuits are configured as two groups:
one MPI rank runs on each node for circuits that have less than 32 qubits, and 2 and 4 MPI ranks on a node (i.e., 512 and 1024 MPI ranks on 256 nodes in total)
for circuits with 35-37 qubits.
(MPI ranks are also referred as \emph{Cores} in the figures).
Figure~\ref{fig:improve} shows the simulation performance of \svsim compared to \intel for each circuit.
The comparison results are presented as the improvement factor (referred as factor below) normalized over the simulation performance of \intel using the same resources.
Any value above 1 means the specific approach improves the total runtime over \intel.

Overall, \svsim outperforms the \intel for all the circuits via \dagp strategy on various numbers of MPI ranks, where the improvement factor for the maximum end-to-end execution time ranges from $1.15\times$ (\textit{qpe} on 128 MPI ranks) to $3.87\times$ (\textit{adder} on 1024 MPI ranks),
with a geometric mean of $1.7\times$ across all the MPI rank configurations. 
%
The right subfigures in Fig.~\ref{fig:improve} show circuits with larger number of qubits, evaluated on 512 and 1024 cores. As the number of qubits and the computational resources increase,
our algorithms scale better compared to \intel baseline.
bv, ising, and cc graphs with 30 qubits have improvement factors up to
$1.7\times$, $1.9 \times$, and $2 \times$ whereas
bv35 has up to $2.7\times$, ising35 has up to $2.5 \times$, and cc36 has up to $3.2\times$.
This demonstrates the benefit of introducing circuit partitioning and hierarchical simulation. 

Comparing across three partitioning strategies, \dagP consistently results in fastest simulation time for all the circuits but qpe.
Take the 256 MPI ranks cases for example, the \dagp offers 29\% and 30\% higher improvement factors over \nat and \dfs on average. In addition, at the 128 and 256 MPI ranks configurations, the improvement factors obtained with \nat and \dfs partitioning show a decreasing trend (7 of 10 circuits) compared to their 32 and 64 ranks configurations, while \dagp exhibits the opposite trend indicating that \dagp partitioning provides a consistent improvement over the baseline simulator.

In summary, \dagp partitioning offers a mean of 2.1 improvement over the baseline \intel simulation
results across all 13 circuits when largest number of MPI ranks applied (i.e. 256 and 1024 for the corresponding circuits).
We demonstrate that the simulation for the circuits with a larger number of qubits
($\geq 35$) have more prominent improvement factors (from $2.5\times$ to $3.9\times$ with the average $3.0\times$).
This shows the effectiveness of \dagp partitioning approach over the baseline and the other two partitioning strategies.

\subsection{Strong scaling}

\begin{figure*}[h!]
     \centering
     \begin{subfigure}[b]{0.16\textwidth}
         \includegraphics[width=\textwidth]{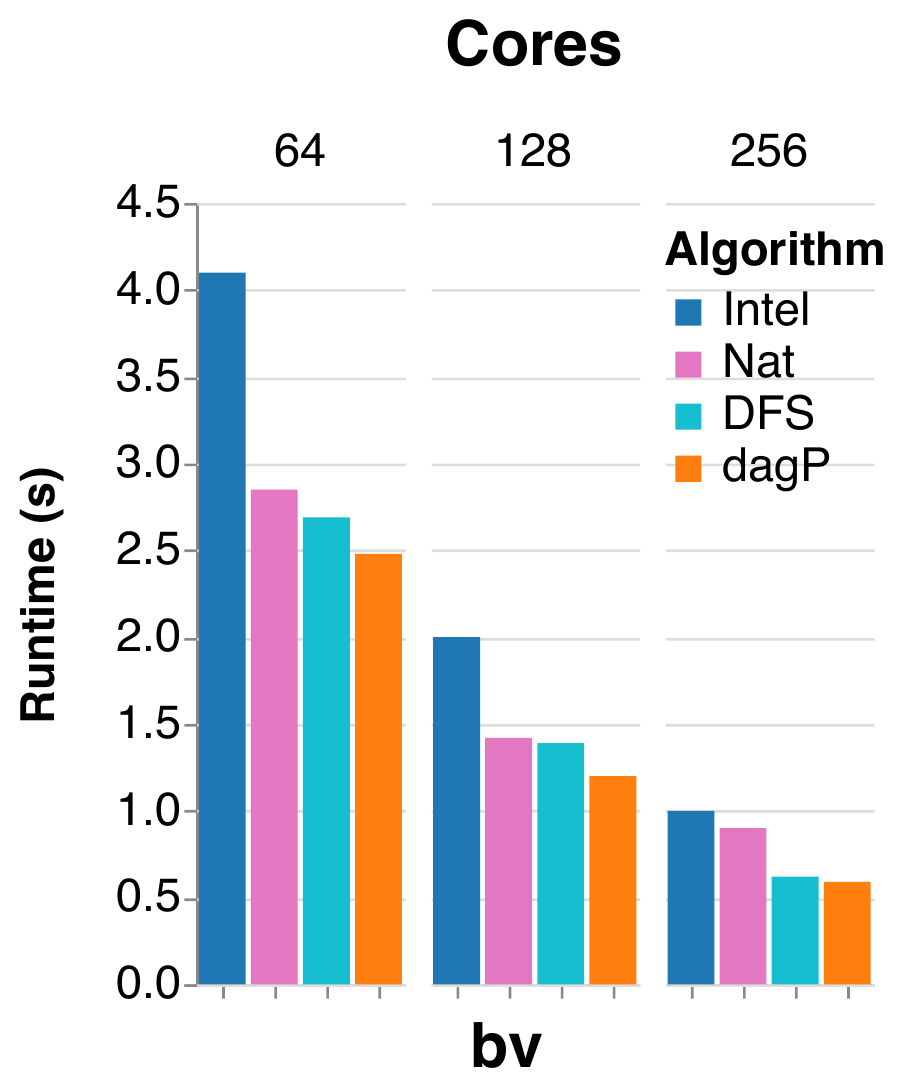}
        \label{fig:bv_speedup}
     \end{subfigure}
    \begin{subfigure}[b]{0.16\textwidth}
         \includegraphics[width=\textwidth]{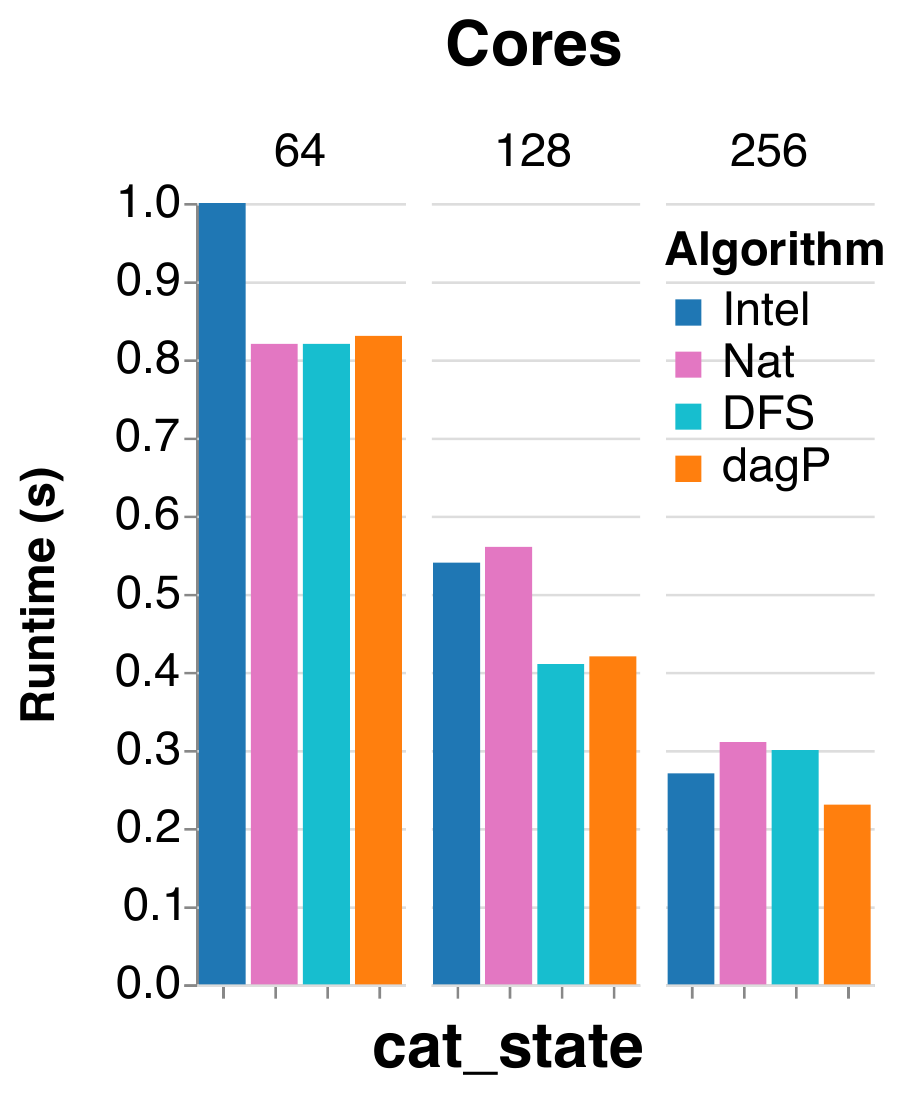}
         \label{fig:cs_speedup}
     \end{subfigure}
     \begin{subfigure}[b]{0.15\textwidth}
         \includegraphics[width=\textwidth]{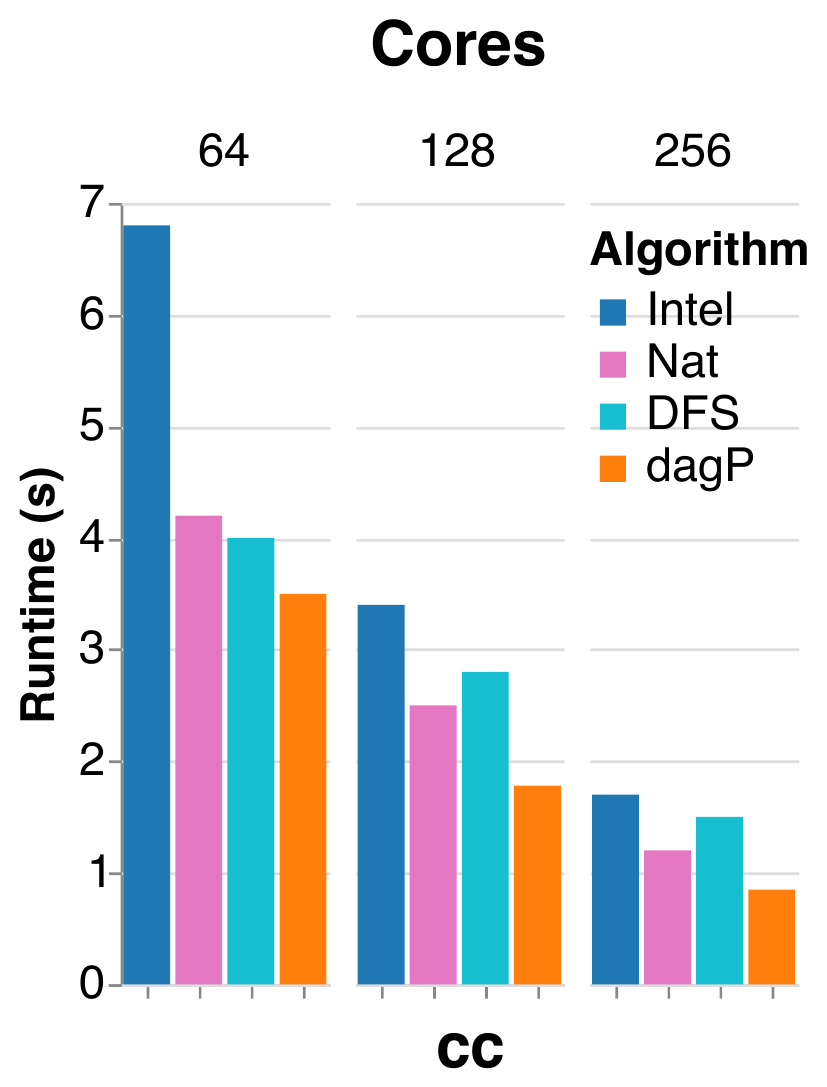}
         \label{fig:cc_speedup}
     \end{subfigure}
    \begin{subfigure}[b]{0.155\textwidth}
         \includegraphics[width=\textwidth]{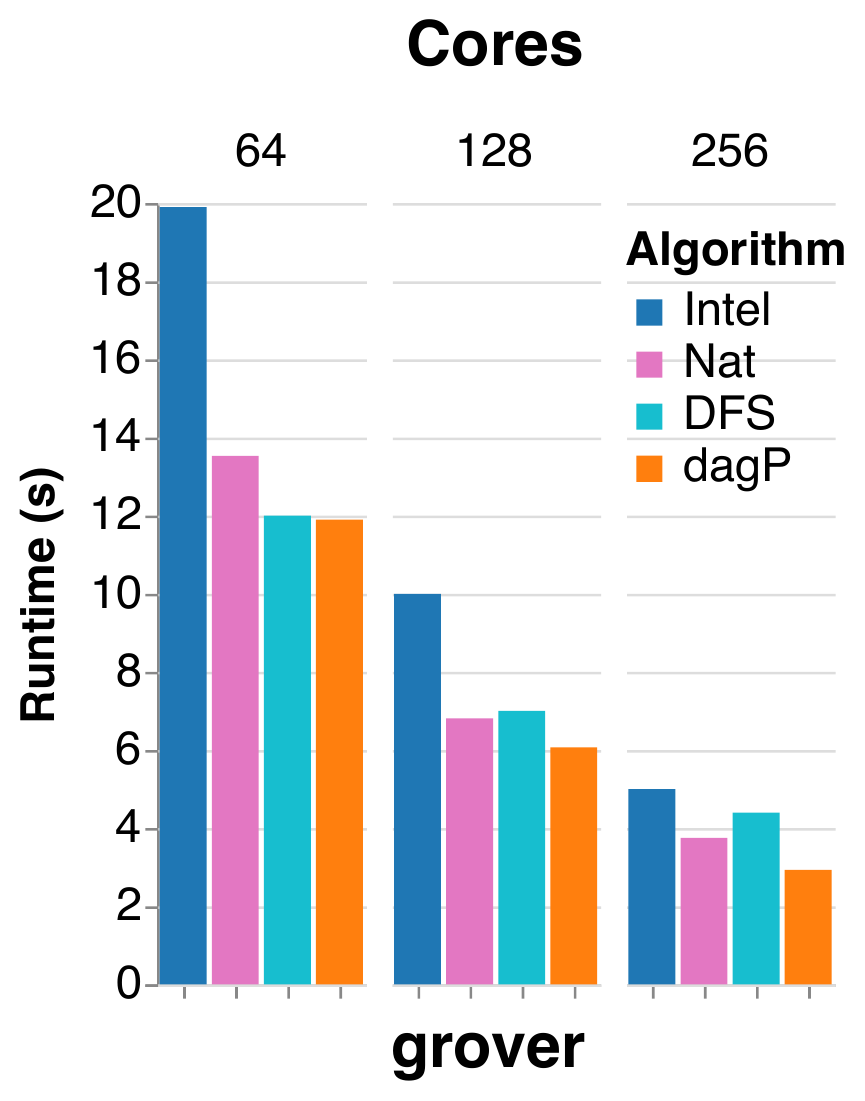}
         \label{fig:grover_speedup}
     \end{subfigure}
     \begin{subfigure}[b]{0.155\textwidth}
         \includegraphics[width=\textwidth]{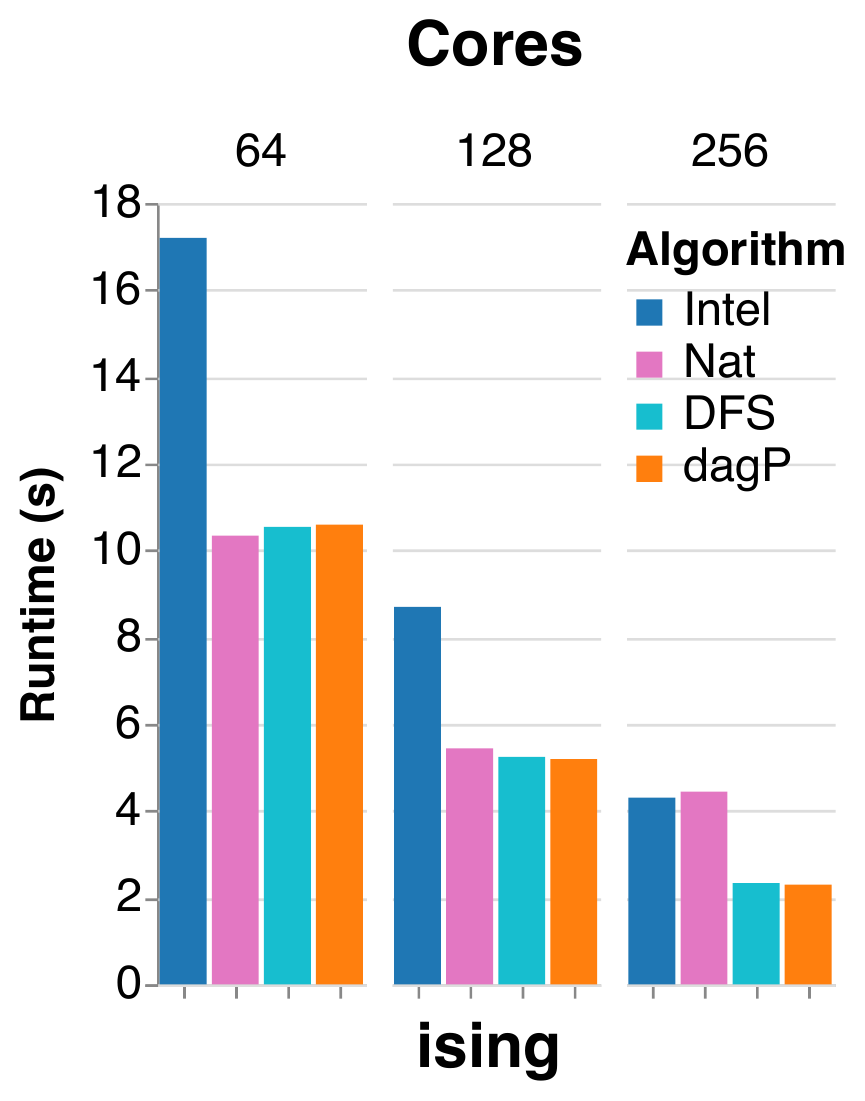}
         \label{fig:ising_speedup}
     \end{subfigure}
     \begin{subfigure}[b]{0.155\textwidth}
         \includegraphics[width=\textwidth]{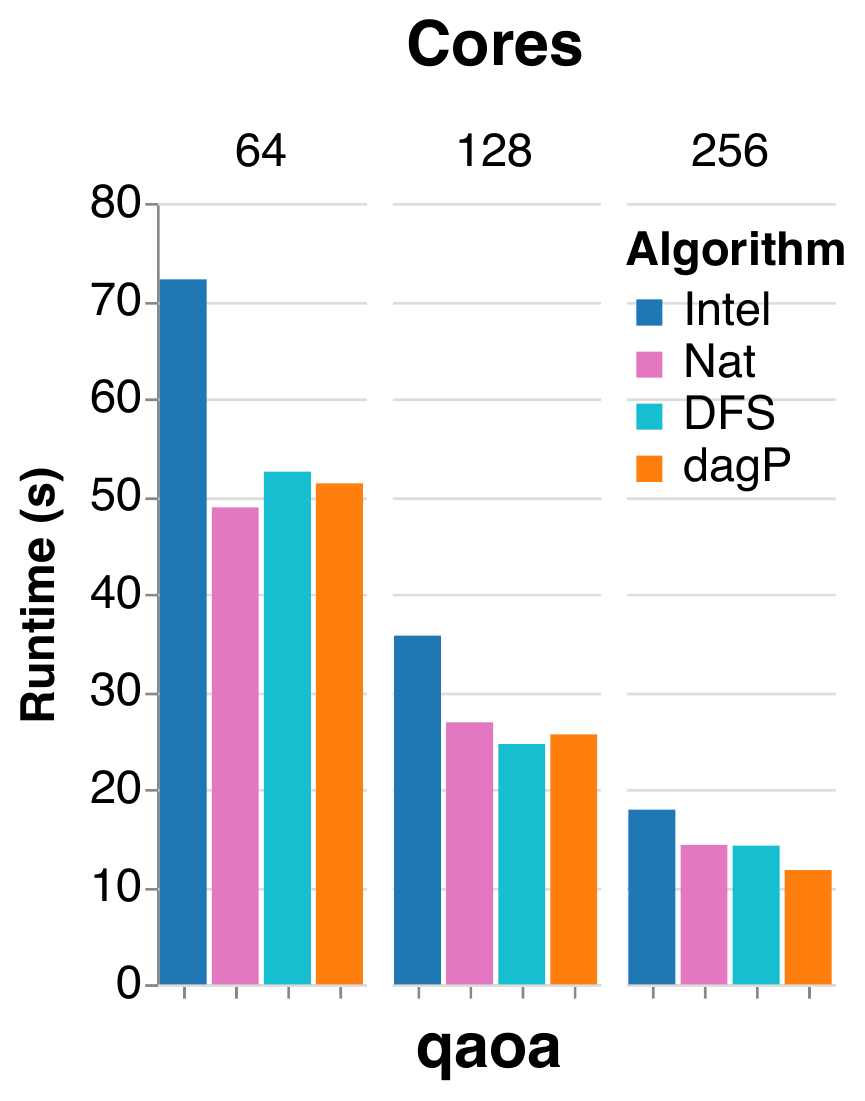}
         \label{fig:qaoa_speedup}
     \end{subfigure}

\vspace{-1em}
    \begin{subfigure}[b]{0.16\textwidth}
         \includegraphics[width=\textwidth]{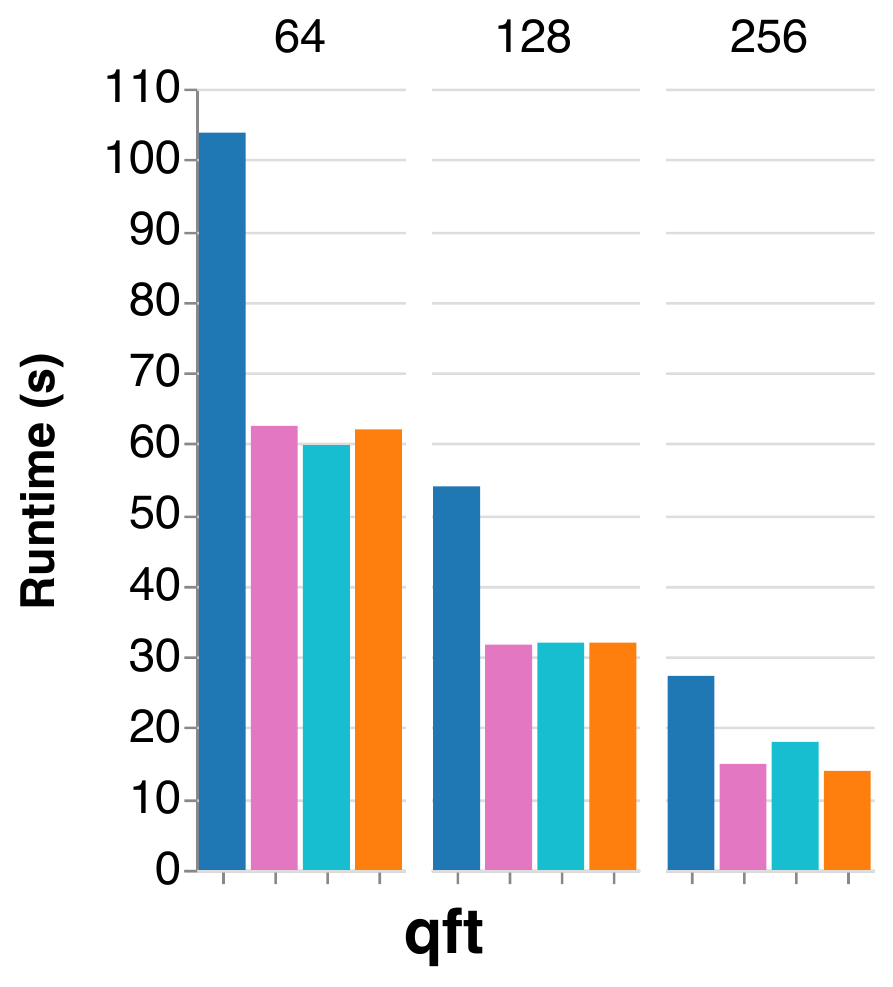}
         \label{fig:qft_speedup}
     \end{subfigure}
      \begin{subfigure}[b]{0.155\textwidth}
         \includegraphics[width=\textwidth]{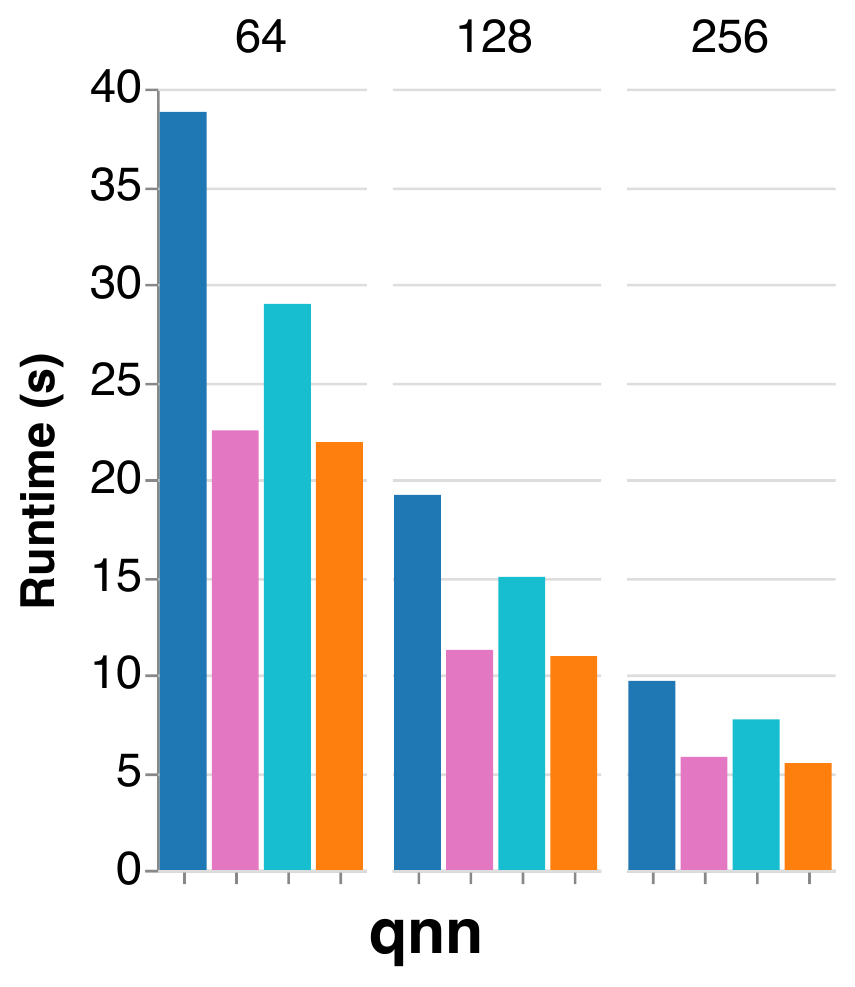}
        \label{fig:qnn_speedup}
     \end{subfigure}
     \begin{subfigure}[b]{0.16\textwidth}
         \includegraphics[width=\textwidth]{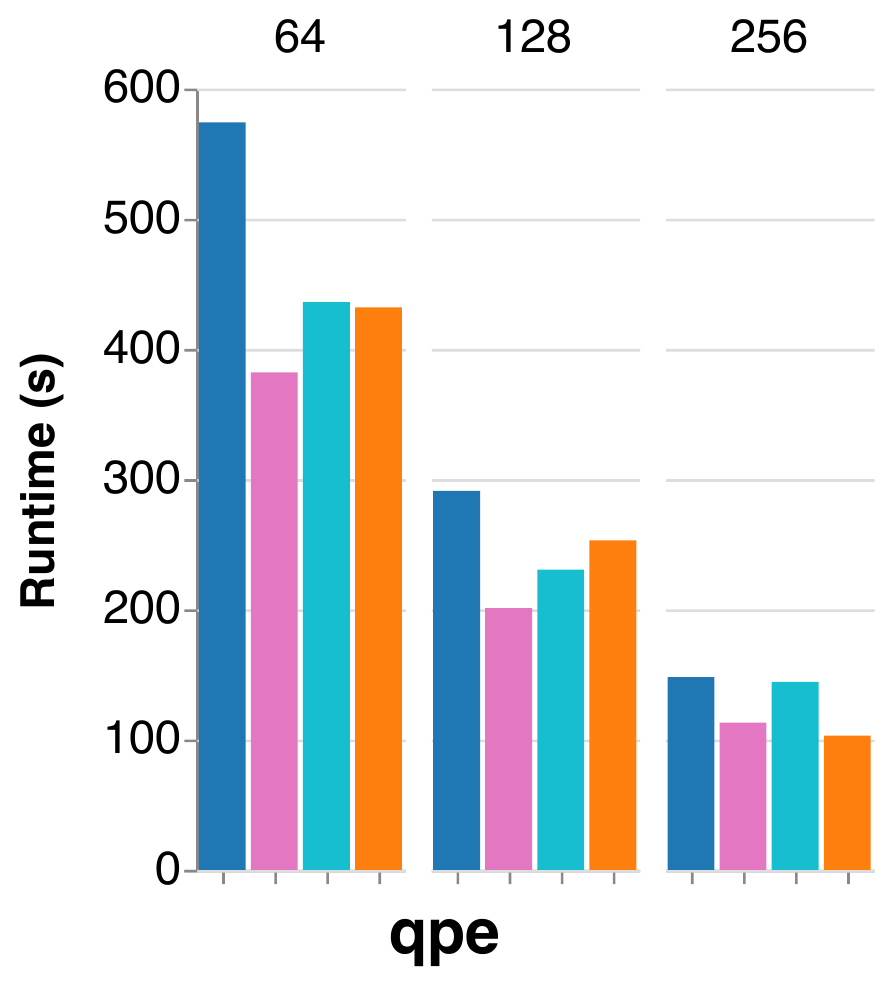}
         \label{fig:qpe_speedup}
     \end{subfigure}
      \begin{subfigure}[b]{0.11\textwidth}
         \includegraphics[width=\textwidth]{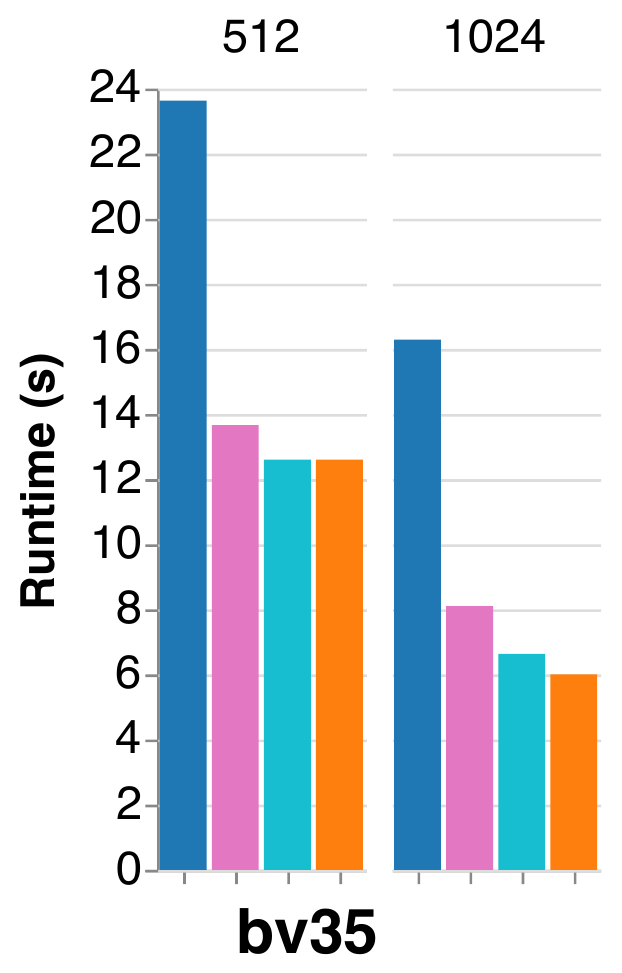}
         \label{fig:bv35_speedup}
     \end{subfigure}
      \begin{subfigure}[b]{0.11\textwidth}
         \includegraphics[width=\textwidth]{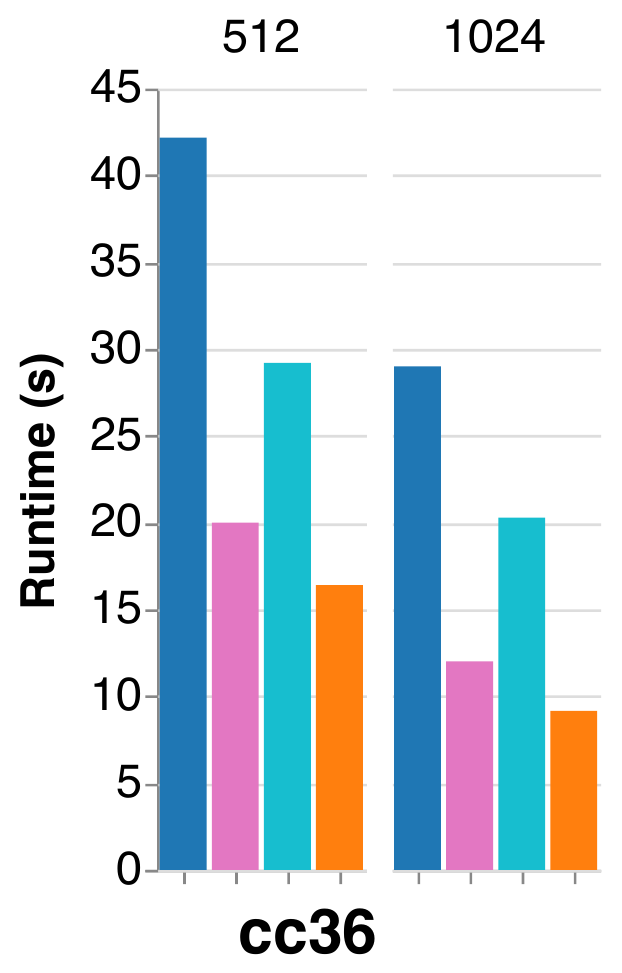}
         \label{fig:cc36_speedup}
     \end{subfigure}
      \begin{subfigure}[b]{0.115\textwidth}
         \includegraphics[width=\textwidth]{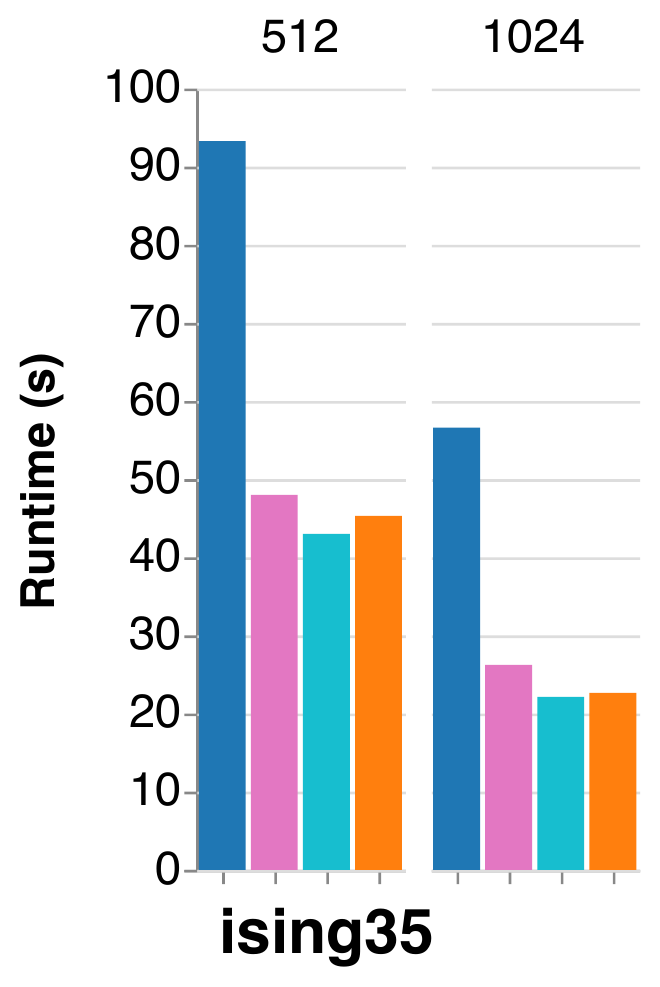}
         \label{fig:ising35_speedup}
     \end{subfigure}
      \begin{subfigure}[b]{0.115\textwidth}
         \includegraphics[width=\textwidth]{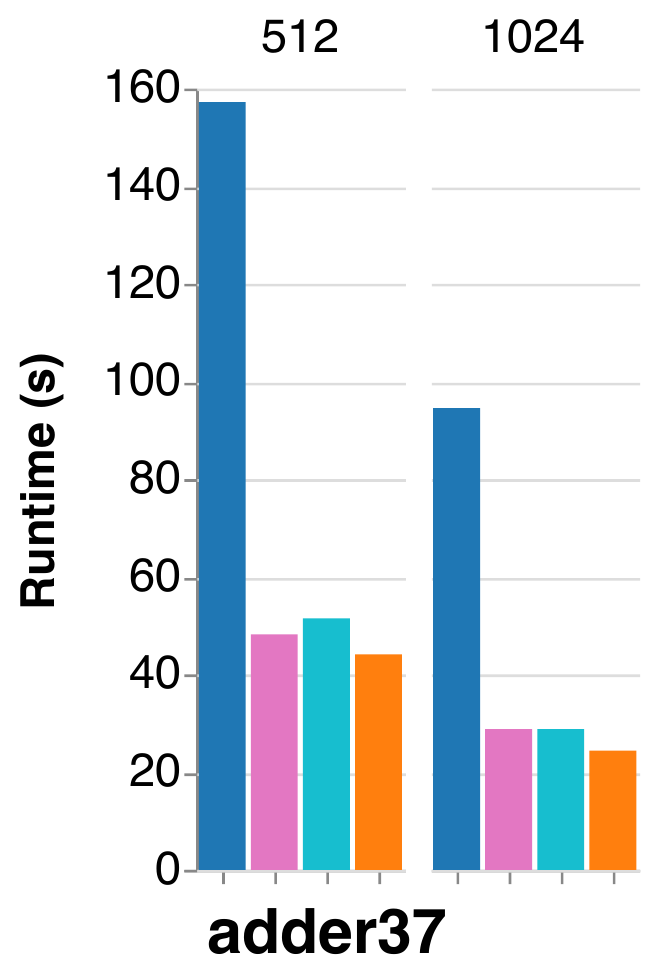}
         \label{fig:adder_speedup}
     \end{subfigure}
     \vspace{-1.5em}
     \caption{Runtime of the input circuits.}
     \label{fig:multinode}
\end{figure*}

\begin{figure*}[h!]
    \centering
    \includegraphics[width=1.01\textwidth]{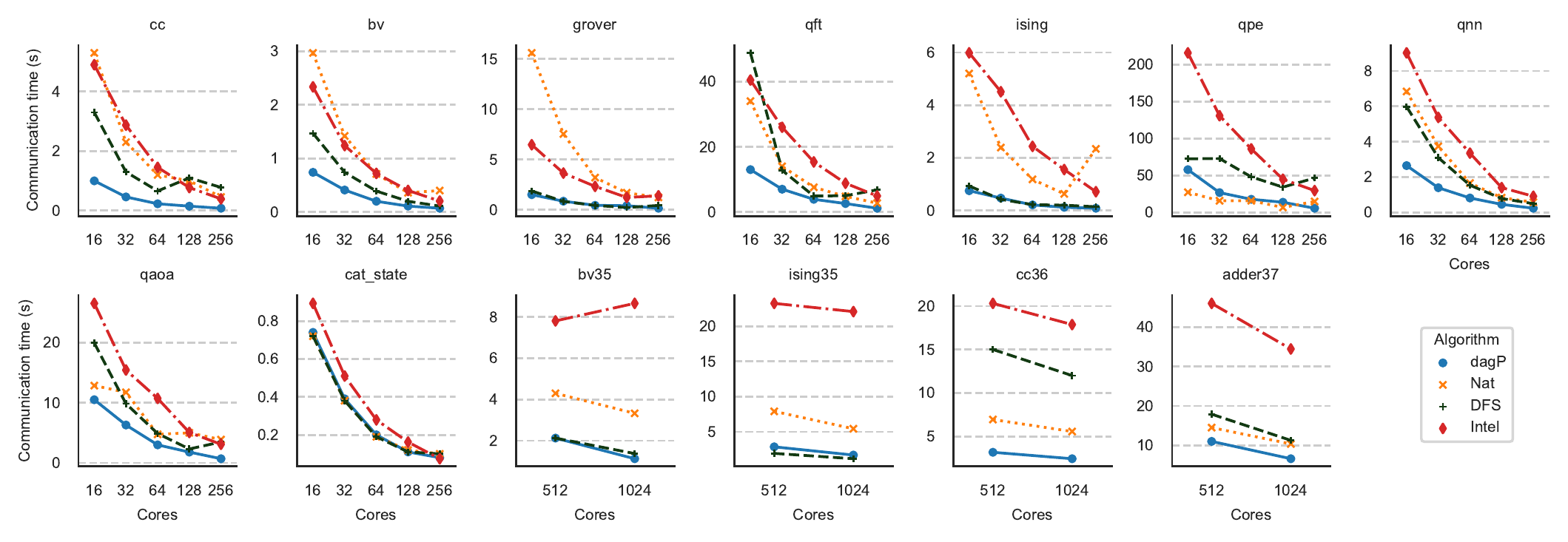}
    \vspace{-1em}
    \caption{Average communication time of the three partitioning variants of \svsim and \intel for all input circuits.}
    \label{fig:nnall_graphs_comm_time_only}
\end{figure*}

In the strong scaling case\footnote{The weak scaling is not considered due to the uniqueness of the quantum circuits: when varying the number of qubits in a circuit, the computation complexity changes non-linearly, and so does the memory footprint and associated data movement complexity.}, Fig.~\ref{fig:multinode} shows the maximum end-to-end simulation time for each circuit partitioned with \nat, \dfs, and \dagp strategies
and the \intel for varying number of MPI ranks. (Due to the limited space, we omit the results for 16 and 32 MPI ranks while the trend is consistent). 
Since \svsim allows computation and communication overlapping, i.e. each rank can continue computation as long as it receives data from other ranks), we report the average MPI communication time across all the ranks per circuit for \svsim. We leverage \textit{remora}~\cite{remora} to obtain the timing details for the computation and MPI calls of \intel.
The minimal overhead of profiling with \textit{remora} is remedied by applying the communication/computation ratio to the end-to-end execution time obtained \emph{without} \textit{remora} present.

We make the following observations:
(I) \svsim shows a close-to-linear speedup for all the partitioning strategies;
(II) For \svsim, both the average computation and communication ratios of the simulation times show
the overall close-to-linear scaling effect across all the configuration cases and circuits;
(III) \svsim consistently offers a faster simulation in the computation portion than the \intel across all circuits.
Note that with \svsim, the average \emph{computation} time is observed similar across different partition strategies.

\begin{figure}
    \centering
    \includegraphics[width=0.48\textwidth,height=12em]{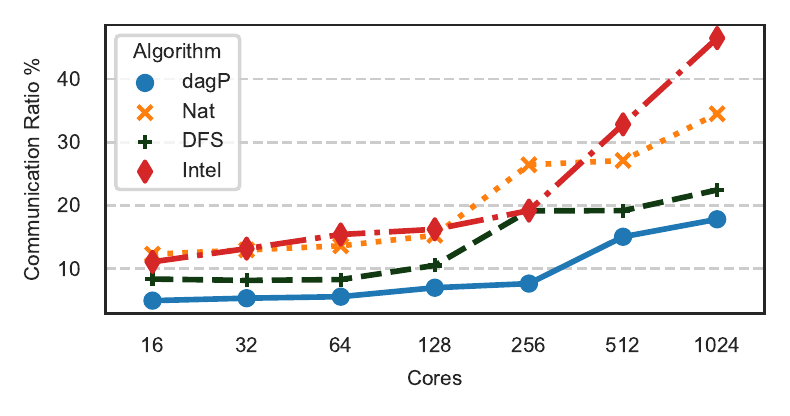} 
    \vspace{-1em}
    \caption{Geometric mean of average communication \textbf{ratio} of the three partitioning variants and \intel for all circuits.}
    \label{fig:nnall_graphs_comm_ratio_only}
\end{figure}

\textbf{MPI communication benefit demonstration.}
Fig.~\ref{fig:nnall_graphs_comm_time_only} shows the per-circuit average communication time for the three strategies and \intel, and Fig.~\ref{fig:nnall_graphs_comm_ratio_only} provides an overall summary.
As seen in the figures, \dagp achieves the fastest communication time across all the cases, and \intel spends relatively longer communication time for all the circuits,
especially for the circuits that contain a larger qubit count.
Fig.~\ref{fig:nnall_graphs_comm_ratio_only} shows that \dagp leads to the lowest geometric mean of the average communication ratio compared to \nat, \dfs and \intel cases for all number of cores tested,
and the \dfs outperforms the \intel except for 256 MPI ranks.
The trend of the lines through increasing number of cores show that \dagp has the best scaling of average communication ratio as well.

\begin{figure}
     \centering
     \begin{subfigure}[b]{0.48\textwidth}
         \includegraphics[width=\textwidth]{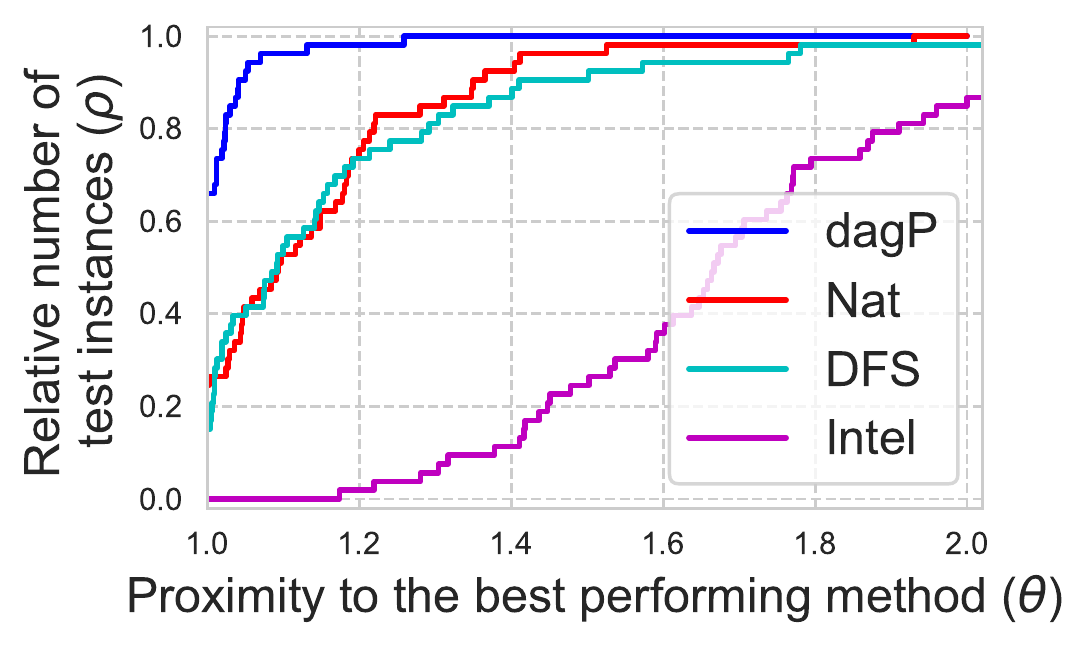}
        \caption{total runtime}
        \label{fig:alg_pp_time}
     \end{subfigure}
     \begin{subfigure}[b]{0.48\textwidth}
         \includegraphics[width=\textwidth]{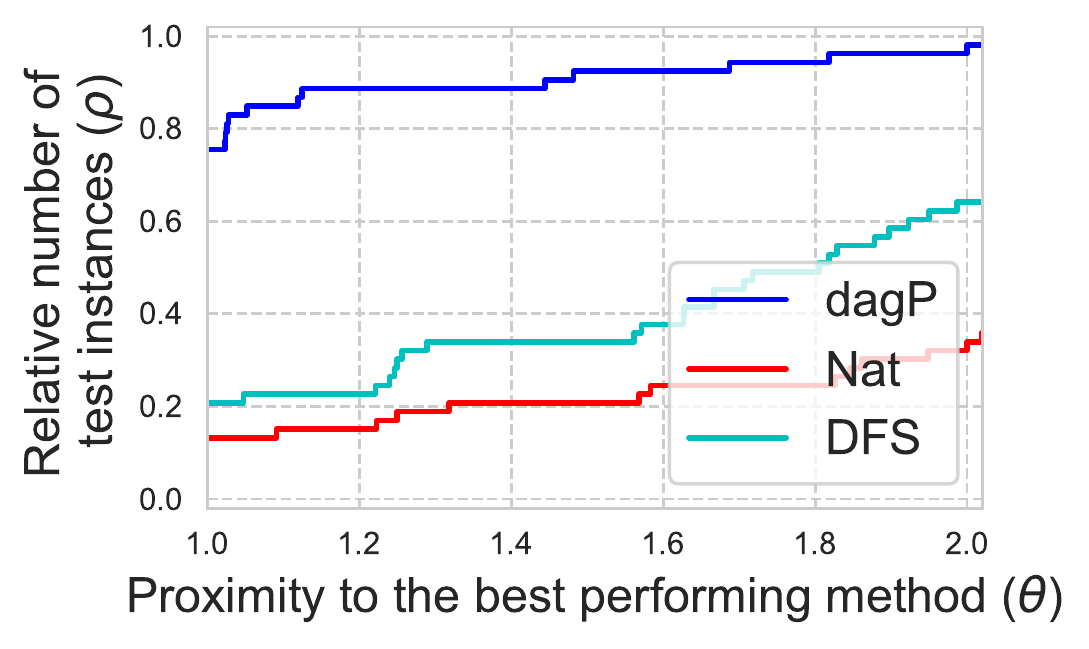}
        \caption{avg. communication time}
        \label{fig:alg_pp_comm_only}
     \end{subfigure}
    \label{fig:prox}
    \caption{Performance profile of time metrics comparison for \intel and three \svsim partitioning methods.}
\end{figure}

Next, for an overall comparison of our algorithm variations, we use performance profiles.
A performance profile shows the ratio ($\rho$) of all input instances where an algorithm performed within a factor ($\theta$) of the \emph{best} performing algorithm~\cite{dolan2002benchmarking}.
Here, an \emph{instance} is a unique test case, e.g., a pair of an input and a number of cores used.
Figures~\ref{fig:alg_pp_time} and~\ref{fig:alg_pp_comm_only} show performance profiles of the three partitioning approaches and  \intel baseline for the total runtime and average communication time, respectively.
For instance, in Fig.~\ref{fig:alg_pp_time}, \dagP performs the best 65\% of the time while \dfs and \nat perform the best approximately 18\% and 25\% of the time.
It also shows \dagP performs within $1.3 \times$ the best total runtime for all instances.
In addition, we see that the \emph{best} result for the \intel is only at $1.2 \times$ the best total runtime.
Similarly, in Fig.~\ref{fig:alg_pp_comm_only}, we see that \dagP has the lowest communication time for 75\% of the input instances.
And, for roughly 40\% of the instances, the other two variants cannot reach an average communication time that is even within $2.0 \times$ the communication time of \dagp approach.

\vspace{-1ex}
\subsection{Multi-level partitioning}
Finally,
we demonstrate the efficiency gained
via exploiting multi-level memory hierarchy in the multi-node system: The first
level partition that uses the main memory of a node to hold the local
state vector and communicates with other nodes to update the local state vector
after finishing a part, and a second level partition that
further partitions the qubits contributing to the local state vector into new
parts, and executes the gates of each new part to improve cache
locality.

In some circuits a natural, intuitive partitioning of the gates
contains less unique qubits than the limits for first-level and second-level partitioning in each part. In this case, the second-level
partitioning returns the identical part to the first-level part at hand. 
Thus, we evaluate the multi-level partitioning on the circuits that contain different sets of parts. For all circuits that
are not shown in the figure, the single- and multi-level \svsim uses the identical partitioning.

\begin{figure}
    \centering
    \includegraphics[width=0.48\textwidth]{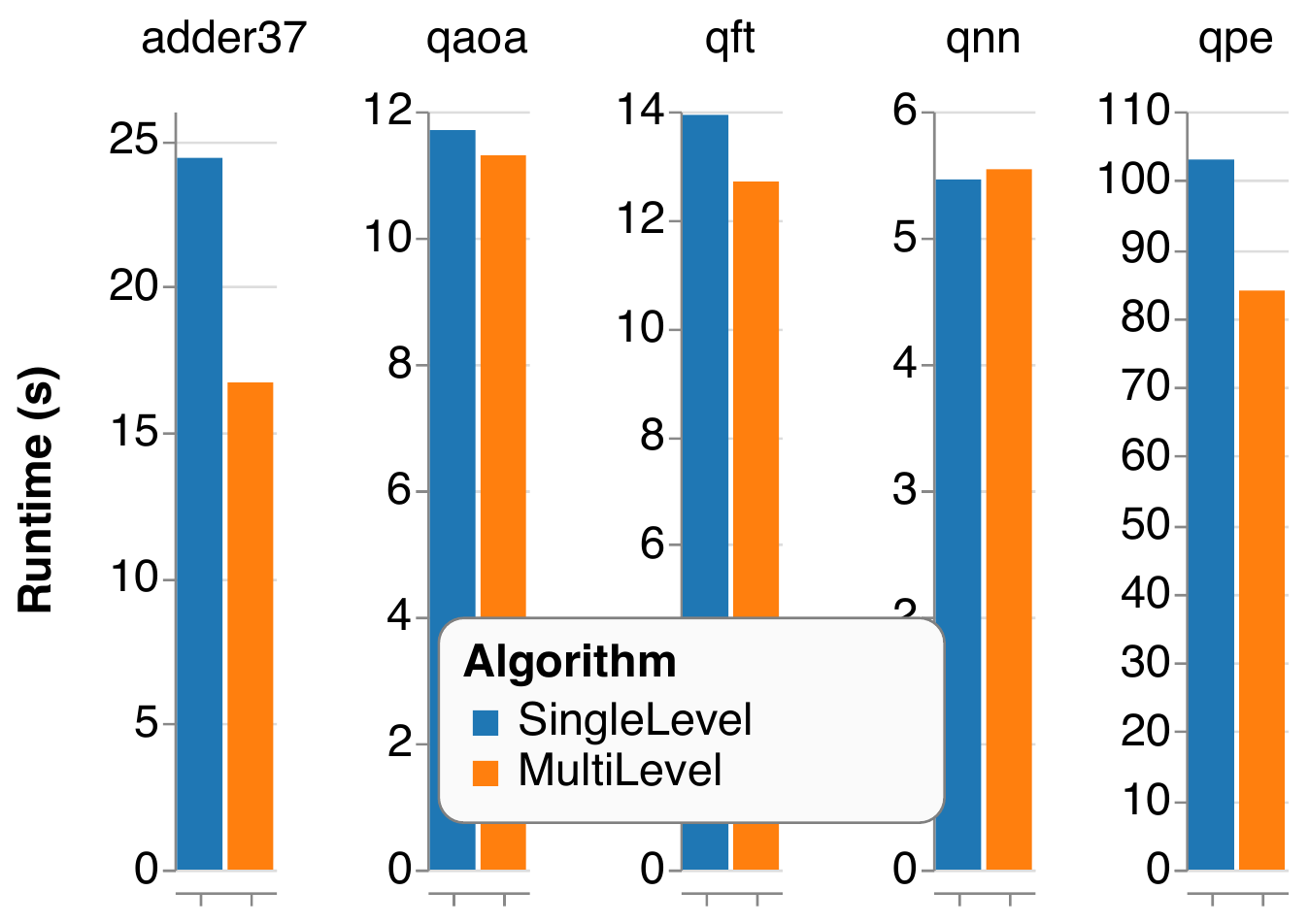}
    \caption{Runtime for the single-level with the best performance and the multi-level.}
    \label{fig:ml_break}
\end{figure}

Figure~\ref{fig:ml_break} shows the execution times of \svsim for the single-level
(i.e., the results shown in Fig.~\ref{fig:multinode})
and multi-level partition with the 256 MPI ranks for qaoa, qft, qnn, and qpe and 1024 MPI ranks for adder.
While the single-level partition results collected represent the most advanced cases by far,
the multi-level partition strategy offers a further improvement over the prior ``best" cases
except for qnn which is 0.1s slower.
The simulation time goes from 24.4s to 16.7s (adder), 14s to 12.7s (qft), 11.8s to 11.3s (qaoa) and 103s to 84s (qpe), with the average of 15.8\% reduction.
This translates to up to $5.67 \times$ improvement over \intel, and $1.47 \times$ over our best single level variant.

\section{GPU Extrapolation}
\label{sec:gpu}

As mentioned earlier, our acyclic graph partitioning enables the optimization space to balance between the local computation and remote communication.
To demonstrate the possible use of our approach with other simulators,
we take the state-of-the-art GPU-based quantum circuit simulator - HyQuas~\cite{zhang2021hyquas} and present a hybrid approach: using HiSVSIM for circuit partitioning and communication and using HyQuas kernel for the computation on GPUs and compare it with the simulation time of using HyQuas on multi-GPU nodes.

We use 4 GPU nodes in our cluster to run the qaoa\_28 circuit (taken from the HyQuas repo) with HyQuas. Each node has an NVIDIA V100-PCIE-16GB GPU. 
Nodes connect via the InfiniBand hardware.
We take the following steps to ensure that the computation to be conducted on GPUs carries the same amount of work as computed by \svsim: \textbf{(i)} We partition the qaoa\_28 circuit into two parts and remap the qubits in each part to model the reordering inside the local state vector. This step ensures the global qubit index is converted to the local slot index. \textbf{(ii)} After remapping, we modify the total qubit number in each part file to fit in the computation model of \svsim.
For example, in four nodes case, to execute a 28 qubit circuit,
\svsim organizes the total qubits as 26 local qubits and 2 MPI process qubits (as described in Sec~\ref{sec:multi}, $log(4) = 2$).
Each part of the original circuit needs to meet the 26 qubits requirement; \textbf{(iii)} we execute the gathered sections of parts (inner state vector) with single-GPU HyQuas on each node.
We replace the local computation component of \svsim (with HyQuas on inner state vector)
and leave the rest of \svsim unchanged.
This way, the cross-node data redistribution stays the same.
Since the whole quantum state fits in the memory of available GPUs,
there is no additional CPU/GPU communication. 


The original qaoa\_28 circuit is partitioned with \dagp, \dfs and \nat as shown in Table~\ref{tab:gpu_part}.
Different strategies generate different number of parts.
The total number of gates matches with the original qaoa\_28 circuit in the HyQuas paper.
Table~\ref{tab:gpu_part} shows the execution time of each part executed with single-GPU HyQuas on one V100 GPU per node.
One can observe that the total execution time on GPU for different strategies is close to each other, which is similar to the \svsim multi-node CPU cases.

Table~\ref{tab:gpu_comp} shows the end-to-end performance estimate on this hybrid approach - using \svsim for partitioning and communication and HyQuas for computation. The result shows that the hybrid approach with \dagp outperforms original HyQuas, indicating that the acyclic partitioning could accelerate data communication for a highly-optimized GPU-based simulator like HyQuas.
Also, among the three strategies, \dagp outperforms \nat and \dfs, which shows promise for future \svsim-\dagp GPU implementation. 




\begin{table}[]
\centering
\caption{QAOA Partitioning breakdown and runtimes on GPUs. qubits: the number of qubits in each part.}
\begin{tabular}{lc@{}r@{}rrrrr}
\toprule
\textbf{Strategy} & \textbf{parts} & & \textbf{qubits} & \textbf{gates} & \textbf{total gates} & \textbf{time (ms)} & \textbf{total (ms)} \\
\toprule
\multirow{2}{*}{\textbf{\dagP}} & \multirow{2}{*}{\textbf{2}} & $P_0$ & 22 & 747 & \multirow{2}{*}{= 1652} & 146.1 & \multirow{2}{*}{\textbf{329.8}} \\
    & & $P_1$ &   24 & 905 &  & 183.7 & \\ \midrule
\multirow{3}{*}{\textbf{\dfs}} & \multirow{3}{*}{3} & $P_0$ & 24 & 536 & \multirow{3}{*}{= 1652} & 110.8 & \multirow{3}{*}{337.7} \\
 & & $P_1$ & 24 & 637 & & 129.7 & \\
 & & $P_2$ & 20 & 479 & & 97.1 & \\ \midrule
\multirow{6}{*}{\textbf{\nat}} & \multirow{6}{*}{6} & $P_0$ & 24 & 24 & \multirow{6}{*}{= 1652} & 11.9 & \multirow{6}{*}{365.9} \\
 & & $P_1$ & 24 & 474 & & 97.7 & \\
 & & $P_2$ & 24 & 260 & & 64.3 & \\
 & & $P_3$ & 24 & 305 & & 65.9 & \\
 & & $P_4$ & 24 & 307 & & 65.8 & \\
 & & $P_5$ & 17 & 282 & & 60.1 & \\
\bottomrule
\end{tabular}
\label{tab:gpu_part}
\end{table}

\begin{table}[t]
\centering
\caption{Estimated QAOA circuit simulation times combining \svsim and HyQuas.}
\begin{tabular}{l@{}rrr}
\toprule
\textbf{Strategy} & \textbf{Communication (s)} & \textbf{Computation (s)} & \textbf{Total time (s)} \\
\toprule
\textbf{\dagP} & \textbf{0.5} & \textbf{0.33} & \textbf{0.83} \\
\textbf{\dfs} & 1.0 & 0.34 & 1.34 \\
\textbf{\nat} & 2.4 & 0.37 & 2.77 \\ \midrule
\textbf{\texttt{HyQuas}~\cite{zhang2021hyquas}} & - & - & 1.47 \\
\bottomrule
\end{tabular}
\label{tab:gpu_comp}
\end{table}

\section{Conclusion}
\label{sec:conclusion}

We present a novel multi-level, distributed hierarchical state vector simulator
\svsim that employs the graph partitioning algorithms for efficient circuit
simulation. The graph partitioning algorithm includes the
acyclic-partitioning-based computation ordering approach.
We evaluate the efficiency of \svsim with various well-known and representative
quantum circuits on Frontera supercomputer with up to 256 nodes and 1024 cores and GPU computation with Nvidia V100.
The results show that \svsim, both multi-level and single-level, scales well and
achieves a significant improvement over the state-of-the-art open-source
distributed quantum circuit simulation systems. The proposed graph-based approach can
be useful for other accelerator-based (i.e., GPUs) quantum simulators and real
quantum computing platforms.

\section*{Acknowledgement}
This material is based upon work supported by the U.S. Department of Energy, Office of Science, National Quantum Information Science Research Centers, Quantum Science Center. The Pacific Northwest National Laboratory is operated by Battelle for the U.S. Department of Energy under Contract DE-AC05-76RL01830.

\bibliographystyle{IEEEtran}
\bibliography{citations}

\end{document}